\documentclass[
reprint,
groupedaddress,
amsmath,amssymb,
aps,
prb,
showkeys,
]{revtex4-2}

\usepackage{graphicx}
\usepackage{dcolumn}
\usepackage{bm}

\usepackage{slashed}
\usepackage{subcaption}

\usepackage[colorlinks=true,
            linkcolor=blue,
            urlcolor=blue,
            citecolor=blue]{hyperref}

\newcommand{\bfq}{\mathbf{q}}
\newcommand{\bfp}{\mathbf{p}}
\newcommand{\nbfq}{|\mathbf{q}|}
\newcommand{\bfx}{\mathbf{x}}
\newcommand{\nbfx}{|\mathbf{x}|}
\newcommand{\calj}{\mathcal J}
\newcommand{\caln}{\mathfrak{n}}
\newcommand{\calm}{\mathfrak{m}}
\newcommand{\call}{\mathfrak{l}}
\newcommand{\frakj}{\mathfrak{j}}
\newcommand{\twop}{(2 \pi)}

\newcommand{\bfs}{\boldsymbol{\sigma}}

\def \a {\alpha}
\def \b {\beta}
\def \g {\gamma}

\def \e {\epsilon}
\def \k {\kappa}
\def \l {\lambda}
\def \m {\mu}
\def \n {\nu}
\def \r {\rho}
\def \s {\sigma}
\def \ra {\rangle}
\def \la {\langle} 
\def \nn {\nonumber}

\begin{document}


\title{Many-body and QED effects in electron-atom inelastic scattering in electron energy loss spectroscopy}

\author{Ioannis Iatrakis}
 \email{ioannis.iatrakis@thermofisher.com}
\affiliation{Materials and Structural Analysis Division, ThermoFisher Scientific,
Achtseweg Noord 5, 5651 GG, Eindhoven, The Netherlands}%


\author{Valerii Brudanin}
 \email{v.s.brudanin@tudelft.nl}
\affiliation{
Department of Imaging Physics, Delft University of Technology, Mekelweg 5,
2628 CD, Delft, The Netherlands
}%

\date{\today}
\begin{abstract}
The elemental composition and electronic structure of materials analyzed by electron energy loss spectroscopy (EELS) are probed by the inner-shell ionization of atoms. This is a localized process that can be approximated by the scattering of an electron beam from a free atom. We calculate the inelastic differential cross section perturbatively within the framework of quantum electrodynamics (QED). The interaction between the incoming electron and the atom factorizes into a high-energy electron term and the atomic transition current. The matrix elements of the transition current are computed within the relaxed Dirac Hartree Fock method. We analyze the correlation effects arising from the relaxation of the atomic orbitals induced by the creation of a core hole. These effects are particularly relevant in quantum many-body systems and have a significant impact on the shape of the differential cross section near the ionization threshold in EELS spectra. In addition to the continuum, we calculate the discrete excitation spectrum of $\mathrm{DyScO_3}$ using crystal-field multiplet theory. The calculated spectrum shows very good agreement with experimental EELS data.
\end{abstract}

\maketitle
\newpage


\section{\label{sec:Intro}Introduction}
The inelastic scattering of fast charged particles with heavy targets is a primary process for studying the structure of matter at different energy scales \cite{Drell:1963ej, PhysRevD.110.056032}. The electron-matter interaction is utilized to reveal electronic structure information of materials in transmission electron microscopy (TEM) through energy dispersive x-ray (EDS) and electron energy loss  spectroscopy (EELS) \cite{doi:10.1126/science.1148820}. To derive the elemental composition of a material from EELS data, the shape of the continuum part of the ionization edges is usually modeled within the approximation of a free, static atom \cite{Rez}.
 
The primary electron energy in TEM is usually within the relativistic regime. Hence, the scattering of the beam from the sample can be described perturbatively as a series in $\alpha = e^2/ \hbar c $ within quantum electrodynamics (QED) \cite{Drell:1963ej}. There exist several calculations of the inelastic differential cross section of an electron with an atom aiming at core loss EELS data analysis, in the context of nonrelativistic quantum mechanics \cite{PhysRevA.6.1013, Rez}. Those have been extended with a semiclassical relativistic approximation of the electron-target interaction \cite{Schattschneider2005}. Similar approaches consider only the projection of the transition current along the beam direction, as it is the main contribution in the case of low-momentum transfer \cite{PhysRev.102.385, PhysRevB.74.064106, sorini2008, 10.21468/SciPostPhys.10.2.031}. In another work, the Breit correction was added to the Coulomb interaction of the beam with the target, which is the first term in the expansion of the scattering matrix in $v_i/c$, where $v_i$ is the velocity of the incident electrons \cite{PhysRevA.77.042701}. The total ionization cross section of the $K$ and $L$ shell electrons in the relativistic framework with simplified calculations of the atomic structure,  based on the central Hartree-Slater one-particle potential, was reported in \cite{PhysRevA.18.963}. 

An important aspect of the cross-section calculation is the electromagnetic current matrix elements of the target atom. In many ionization cross-section calculations for EELS, the atomic state is calculated within a mean-field, one-particle approximation. The atomic state is also assumed to be frozen during the scattering event, and the core-hole impact on the final atomic state is not included. Then, the asymptotic form of the one-particle mean-field potential is typically adjusted to match the expected large-distance behavior of the cross section \cite{PhysRevA.6.1013, Rez, PhysRevA.18.963, Zhang_2025}. In other cases, the atomic states are calculated in effective one-particle density functional theory (DFT) potentials \cite{segger_2023_7645765}. However, single-particle potentials for the correlation effects of inner atomic shells are only approximated by considering semi-local information in density functionals \cite{PhysRevB.23.5048, PhysRevB.47.15404, MYaAmusia_2003}.

The Dirac-Hartree-Fock (DHF) method has yielded high-accuracy results for both structural and dynamic properties of atoms, without relying on an approximate one-particle potential for the exchange interaction \cite{DESCLAUX197531, Desclaux:1973qgc, indelicato_lindroth_1992, JONSSON2007597, PhysRevA.94.012502, ambit, PhysRevA.54.3948}. Further improvements can be achieved by incorporating correlation corrections from many-body perturbation theory (MBPT) and configuration interaction (CI) \cite{PhysRevA.54.3948, PhysRev.136.B896, PhysRevA.74.062503}. Relativistic corrections derived from higher-order transverse photon corrections contribute to the description of deep atomic shells especially for heavy atoms \cite{GRANT198837, IPGrant_1974, doi:10.1080/00018737000101191}. The transverse part of the electron-electron interaction is described up to the first order in ${\ O}\left( {v^2 / c^2} \right)$ by the Breit potential \cite{PhysRevA.4.41, indelicato_lindroth_1992}. Further corrections, derived from the one-loop perturbative expansion, have been investigated in the past as well. Those include the Uehling potential, derived from the one-loop vacuum polarization, and the intrinsic one-loop electron self-energy \cite{PhysRevA.72.052115, PhysRevA.93.052509, Ginges_2016}.

In the case of ionization experiments, correlation effects in the final state are important, as has been first addressed in photo-ionization processes \cite{AMUSIA1967541, PhysRevLett.30.529}. These effects are typically attributed to the relaxation of the atomic orbitals in the final state due to the generated vacancy and higher-order contributions from virtual electron-hole pairs, which are described within many-body theory and the random phase approximation with exchange (RPAE) \cite{osti_4466451, Amusia1971, WRJohnson_1980, Rehr_2005, RevModPhys.74.601}.

For inner-shell ionization, the atomic core shells are localized and well separated in energy. Hence,  the diagonal part of RPAE reduces to the self-consistent DHF solution of the ionized atom, with a core hole present and an outgoing ionized electron.  The rearrangement of the atomic core has been shown to have significant influence on the ionization energies and the ionization differential cross-section shapes of atoms, particularly for core-shell electrons \cite{PhysRevLett.30.529,  Amusia_1974, Amusia1975, AMUSIA1976191, Amusia1981, amusia1997inner, refId0, 2006sham.book.379S, indelicato_lindroth_1992}. 


Final-state effects also influence the near edge fine structure of x-ray absorption (XAS) and EELS spectra which is closely related to the unoccupied, projected density of states of the atomic target. The impact of the induced core-hole potential on the empty density of states and excitonic peaks has been studied by employing the final-state rule in DFT calculations \cite{PhysRevB.11.2391, PhysRevB.25.5150, PhysRevLett.85.1298, PhysRevB.41.11899, pantelides1, pantelides2, Shirley}. Still, the local exchange and correlation potential of one-particle DFT needs to be generalized for the description of the near-edge structure of strongly correlated systems to include multiplet effects, nonlocal exchange, charge transfer, Fano mixing and an explicit treatment of the electron-hole exciton \cite{GROOT200531,  HEDIN19701, PhysRevB.64.165112, Rehr_2005, PhysRevB.85.165113, PhysRevB.47.15404}. The electron-hole exciton has been studied within the Bethe-Salpeter equation (BSE), which resembles the static limit of RPAE but it is applied to an extended many-body system instead of an atom. BSE commonly adopts a nonlocal exchange interaction and static screening of the core hole by the electron density of the solid within the GW approximation  \cite{PhysRevB.64.165112, urquiza2024connectionsresonantinelasticxray}. In the context of core-shell excitations, the BSE has been shown to be comparable to the final-state rule with the main differences stemming from the local exchange potentials which are commonly used in the implementation of the final state rule in DFT for large systems \cite{Rehr_2005}. In case of a finite atomic system, the relaxed DHF is explicitly nonlocal so this is a closer approximation to BSE in the limit where intershell and intrashell correlations are small. Non-static generalization of the electron-hole interaction goes beyond the relaxed DHF to the full RPAE which has been applied to outer-shell photoionization and ground-state properties of single atoms, \cite{Amusia_1974, PhysRevA.74.062503}.

Dynamic intershell and intrashell correlation effects play an important role in ionization of valence and subvalence shells. Such effects have been described within  RPAE,  MBPT, and CI including orbital relaxation due to the vacancy. Comparisons among the different methods have been performed for outer shell ionization of free atoms such as Ar, Ne, and Xe in \cite{PhysRevA.23.2394, PhysRevA.37.4671, PhysRevLett.30.529}. These higher-order correlation effects are small for inner-shell atomic properties as it has been shown in \cite{PhysRevA.31.556, ELindroth_1993} and we do not focus on them in this work.

Our work aims to apply the generic framework of inclusive inelastic scattering within quantum field theory to EELS \cite{Drell:1963ej}. The differential cross section factorizes to the high-energy electron kinematic part and the target structure tensor, which encodes the information on the charge and current density of the target. The factorization of the cross section is a generic result which has been discussed in the context of inclusive inelastic scattering of electrons with nuclear targets \cite{DeForest:1966ycn}. We are primarily interested in describing core-loss EELS spectra, which arise from inner-shell ionization of atoms. Given the spherical symmetry of the atomic core, the transition 4-current is expanded in terms of multipole operators, classified by their transformation properties under $\mathrm{SO(3)}$. This multipole decomposition enables the identification and separation of the longitudinal and transverse transition channels. We derive the angular dependence of the transition matrix analytically in the basis of angular momentum eigenstates for a generic relativistic target. The cross section is then formulated in terms of one-dimensional radial integrals over the orbital wave functions. The result applies up to very high energy loss and momentum transfer. We calculate the long-wavelength limit via a systematic expansion at low-energy loss and momentum transfer.

Focusing on the analysis of core-loss EELS ionization edges, we calculate the atomic structure within the relaxed DHF framework, using the {\it ab initio}, high-accuracy atomic structure software \texttt{AMBIT} \cite{ambit}. Hence, we employ the full nonlocal exchange interaction in the self-consistent solution. The final atomic state following an inner-shell ionization is computed self-consistently by explicitly including the core hole, which induces the relaxation of the atomic core. We emphasize the equivalence of the self-consistent relaxed final state to the time-forward diagrams in RPAE, by a perturbative expansion of the final state in terms of the core-hole potential. We analyze in detail how the ionization differential cross-section shape depends on the presence of the core hole and how it compares to the frozen core and $Z+1$ approximations. A final state with a decayed core hole is also considered in our quantitative comparison of different final-state approaches. 

EELS is also utilized for the study of the excitation spectra of transition metal and rare-earth elements \cite{GROOT200531}. The excitonic spectrum of these elements reveals information about their oxidation state and the electronic structure of the valence band.  We calculate the total excitation and differential ionization cross section for scandium $L$  and dysprosium $M$ edges, using crystal-field multiplet theory \cite{PhysRevB.32.5107, PhysRevB.41.928}. The inclusion of the core hole in the final-state Hamiltonian of the system is known to be important to the spectral shape \cite{ PhysRevB.41.928}.

The structure of this article is as follows. In Sec. \ref{sec:qed}, we derive the general expression of the inelastic differential cross section of an electron with a spherically symmetric heavy target in terms of the reduced matrix elements of the transition current of the target. The result is also compared with that of the multipole photoionization cross section. In Sec. \ref{sec:atomicstructure}, the reduced matrix elements of the transition current are calculated for a relativistic target, which is described as a many-particle state in the basis of Dirac orbital wave functions. The low-momentum transfer limit is found in Sec. \ref{subsec:lowq}. The Hamiltonian of the target atoms is introduced in the DHF scheme. We then discuss the impact of the core hole on the calculation of the final state after the electron-atom collision. In Sec. \ref{sec:results}, we present a detailed discussion of our numerical results, and compare different approximations. We also calculate the total excitation and ionization spectrum of $\textrm{DyScO}_3$  and compare it with experimental EELS data. We finally summarize our results and discuss future directions in Sec. \ref{sec:sum}.

\section{\label{sec:qed}Inelastic electron scattering}

\subsection{Quantum electrodynamics}
The electron interaction with a heavy target is described by the minimal coupling of an external electromagnetic current operator $\calj^\mu(x) = (\rho(x), {\boldsymbol \calj}(x))$ to the photon field in the QED Lagrangian
\begin{eqnarray}
    {\cal L} =  - \frac{1}{4} F^{\m\n} F_{\m\n}
- \overline{\Psi} \left( [\slashed{\partial} +i e \slashed{A} ]+ m \right) \Psi + \calj^\m A_\m
 \, ,
\label{QEDLagrangian}
\end{eqnarray}
where $A_\mu$ is the photon field, $F_{\mu\nu} = \partial_\mu A_\n - \partial_n A_\m$, and $\Psi$ is the electron spinor field. The metric convention is mostly positive, $\eta_{\mu\nu} = \textrm{diag}(-1,1,1,1)$, and the gamma matrices $\gamma^\mu$ follow the Weyl notation.The standard notation is adopted, $\slashed{A} = \gamma^\m A_\m$ and $\partial_\m = \frac{\partial}{\partial x_\mu}$. The electromagnetic current operator is conserved, $\partial_\mu {\mathcal J}^\m(x) =0$. We adopt the system of natural units, where $\hbar = c = 1$.

\begin{figure}[ht]
\includegraphics[width=.4\linewidth]{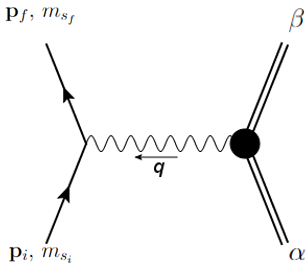}
\caption{\label{fig:treelevel} The tree level diagram for the inelastic scattering of an incoming electron ($| i \rangle \to | f \rangle$) from a heavy target which is excited from the initial multiparticle state $| \Phi_\alpha \rangle$ to the final $| \Phi_\beta \rangle$. In our case $| \Phi_\beta \rangle $ will be either an ionized or an excited state.}
\label{feynmandiag}
\end{figure}

The inelastic scattering of high-energy electrons from a heavy target that interacts electromagnetically is described at tree level by the Feynman diagram in Fig. \ref{feynmandiag}. 
The S-matrix of the scattering from the initial state $ | I \rangle = |p_i, m_{s_i} \rangle \otimes | \Phi_\a \rangle$ to the final $ | F \rangle = |p_i, m_{s_i} \rangle \otimes | \Phi_\beta \rangle$  in momentum space reads as
\begin{align}
S_{F;I} &=
\langle p_f,s_f,\,\Phi_\beta | S | p_i,s_i,\,\Phi_\alpha \rangle
\nonumber \\ &
=\frac{e}{(2 \pi)^2} \delta(E_i-E_f-E_\b+E_\a) 
\nonumber \\  & \times
\frac{\bar{u}(p_f, m_{s_f}) \gamma^\m u(p_i, m_{s_i})}{q^2} 
\langle \Phi_\beta | \calj_{\m}(\bfq) | \Phi_\alpha \rangle\,,
\end{align}
where $q = p_f-p_i$ is 4-vector momentum transfer, i.e., $q = (E_f-E_i, \bfp_f - \bfp_i)$. Time translation symmetry implies that the transition current can be written in the form $\calj^\m(q) = e^{i(E_\b- E_\a) t} \int d^3x  e^{- i{\bf q \cdot x}}
\calj^\m({\bf x})=e^{i(E_\b- E_\a) t} \calj^\m(\bfq)$. By factoring out the energy conserving delta function one can retrieve the transition matrix (T matrix) which is defined as
\begin{equation}
S_{F;I} = 
\delta_{F, I}
-2\pi \,i  \delta(E_i-E_f-E_\b+E_\a)
T_{p_f,s_f,\,\beta \,;\, p_i,s_i,\,\alpha}\,.
\label{Sdef}
\end{equation} 
Considering an unpolarized electron beam and the fact that the spin of the beam electrons is not measured in an EELS detector, we average and sum over the initial and final spin states, respectively. The $T$ matrix norm squared is

\begin{align}
|T_{p_f,\,\beta \,;\, p_i,\,\alpha}|^2
= & -\frac{e^2}{2 (2 \pi)^6 q^4} \frac{1}{E_i E_f}
\nn \\
& \times \Big( 
(p_i \cdot p_f +m^2) \eta^{\m\n} 
- p_i^\m\, p_f^\n -p_i^\n \, p_f^\m   \Big)
\nn \\ 
& \times
\langle \Phi_\alpha | \calj_{\n}^{\dagger}(\bfq) | \Phi_\beta \rangle
\langle \Phi_\beta | \calj_{\m}(\bfq) | \Phi_\alpha \rangle
\label{tnorm}
\end{align}
where the beam electron contribution factorizes from the transition current of the target. The detailed derivation is presented in appendix \ref{ap:feynmandiagram}. This is a generic result which is well known from the study of inclusive inelastic scattering in particle physics \cite{Weinberg_1995}. Finally, the inelastic differential cross section of an unpolarized incoming electron beam interacting with a heavy target reads as
\begin{align}    
d \sigma(p_f,\,\beta \,;\, p_i,\,\alpha) = & (2 \pi)^4  \frac{d^3 p_f}{|\mathbf{v}_{i} |}
\sum_\b \int d\b 
\delta( E_i - E_f-E_\b+E_\a) 
\nn \\ &
 \times |T_{p_f,\,\beta \,;\, p_i,\,\alpha}|^2 \,,
\label{dcrosssectionT}
\end{align}
where we sum and integrate over the discrete and continuous quantum numbers of the final state $\beta$. ${\bf v}_i$ is the velocity of the incoming electrons. 


\subsection{\label{subsec:multiexp} Multipole expansion of the T matrix}
The right vertex of the Feynman diagram, Fig. \ref{fig:treelevel}, includes the target transition matrix element and contains the information about the target excitations. We consider that the target's energy spectrum consists of angular-momentum eigenstates $J_\a^2 |\Phi_\a \rangle = J_\a (J_\a+1) | \Phi_\a \rangle$. In this case, we will expand the current matrix element into $\mathrm{SO(3)}$ irreducible tensor operators.

We choose the z axis of our coordinate system along the virtual photon momentum $\bfq$, without loss of generality for an unpolarized sample \cite{DeForest:1966ycn}. For a spherically symmetric target, one can rotate the quantization axis to an arbitrary direction without affecting the final cross section since we sum over all the projections of the angular momentum in the final state.

The 4-current is  expressed in terms of the charge density and the vector transition current $\calj = (\rho, {\boldsymbol \calj})$. The conservation of current in momentum space leads to the  relation of the longitudinal current to the charge density
\begin{equation}
\hat{e}_{(\bfq)} \cdot {\boldsymbol \calj}(\bfq) =\frac{q_0}{\nbfq} \rho(\bfq) \,.
\end{equation}
Equation ~(\ref{tnorm}) can then be written in terms of the charge density and the transverse current

\begin{align}
|T_{p_f,\,\beta \,;\, p_i,\,\alpha}|^2 =&
-\frac{ e^2}{2 (2 \pi)^6 } \frac{1}{q^4} \frac{1}{E_i E_f}
\nn \\
& \times
\left( C_L(q) | \r(\bfq)|^2
+C_T(q) | \boldsymbol{\calj_\perp}(\bfq)|^2
\right)
\end{align}
The coefficients of the longitudinal and transverse terms are
\begin{align}
C_L(q) = {1 \over 2} {q^4 \over \nbfq^4} \left[ \left(E_i+E_f\right)^2-\nbfq^2\right] 
\nonumber \\
C_T(q) = \vert \mathbf{P} \vert^2 - {\left( \bfq \cdot \mathbf{P}\right)^2 \over \nbfq^2} + {q^2 \over 2} \, ,
\end{align}
and ${\mathbf P} = \frac{\mathbf p_i + \mathbf p_f}{2}$. 
The charge and current densities in the coordinate space are
\begin{align}
\r(\bfq) = \int d^3x \, e^{-i \bfq \cdot \bfx} \r(\bfx)
\,,\,\,
{\boldsymbol \calj_\perp}(\bfq) = \int d^3x \,e^{-i \bfq \cdot \bfx} {\boldsymbol \calj_\perp}(\bfx)
\end{align}
The plane-wave expansion in terms of spherical harmonics is applied to the charge density 
\begin{equation}
e^{-i \bfq \cdot \bfx} = \sqrt{4 \pi} \sum_{J} (-i)^J \sqrt{2 J+1} j_J(\nbfq r) Y_J^0(\hat{\bfx}) \, , 
\label{rayleigh}
\end{equation}
where $r = | \bfx |$.
The transverse vector current is expanded in terms of the spherical components of the $\hat{e}_{(\bfq)}$ basis. 
\begin{figure}[b]
\includegraphics[width=8cm]{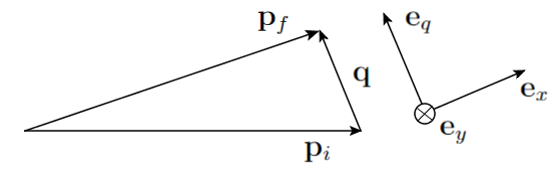}
\caption{The coordinate system of the expansion of the vector current operator is defined with respect to the momentum transfer vector.}
\label{fig:kinematics}
\end{figure}
Then, the $\l$ spherical component of the current is $\calj_\l = \hat{e}_{(\bfq) \, \l} \cdot \calj$. The spherical vector is coupled to the spherical harmonics from the plane-wave expansion [Eq. (\ref{rayleigh})],
\begin{align}
\hat{e}_{(\bfq) \, \l} e^{-i \bfq \cdot \bfx} =& -\sqrt{2 \pi} \sum_{J = 1}^{\infty} (-i)^J \sqrt{2J+1} \Bigg[ \l j_J(\nbfq r)  \mathbf{Y}_{JJ}^\l(\hat{\bfx}) 
\nonumber \\ 
& +i\Bigg(
\sqrt{J+1 \over 2 J +1} j_{J-1} \left(\nbfq \nbfx\right) \mathbf{Y}_{J \, J-1}^\l(\hat{\bfx})
\nn \\ &
-\sqrt{J \over 2 J +1} 
j_{J+1}\left(\nbfq \nbfx\right) \mathbf{Y}_{J \, J+1}^\l(\hat{\bfx})
\Bigg)
\Bigg]
\end{align}
The norm squared of the above expressions is calculated in terms of the irreducible matrix elements of the current operator
\begin{widetext}
\begin{align}
{1 \over 2 J_\a +1} \sum_{M_\a \, M_\b} |T_{p_f,\,\b \,;\, p_i,\,\a}|^2 &={ e^2 \over \twop^5}  {1\over q^4} {1 \over E_i E_f} {1 \over 2 J_\a +1} 
\left[
C_L(q) \, \sum_{J=0}^{\infty} 
\left| \langle \Phi_\b ||T_J^{\mathrm(coul)} || \Phi_\a \rangle  \right|^2 
\right .
\nonumber \\
& \left .
+ C_T(q) \sum_{J=1}^{\infty} 
\left(
\left| \langle \Phi_\b || T_J^{\mathrm(e)} || \Phi_\a \rangle \right|^2  + \left| \langle \Phi_\b || T_J^{\mathrm(m)} || \Phi_\a \rangle \right|^2
\right) \right] \,,
\label{QEDTmatrix}
\end{align}
where we average and sum over the initial and final projections of the angular momentum respectively. The irreducible parts of the current operator under $\mathrm{SO(3)}$ in terms of the vector spherical harmonics are

\begin{align}    
T_J^{M\, \mathrm(coul)}  =& \int d^3 \bfx \, j_J\left(\nbfq \nbfx\right) Y_J^M(\hat{\bfx}) \rho(\bfx) \, ,\,\,\,\,
T_J^{M \, \mathrm(mag)} =  \int d^3 \bfx \, j_J(\nbfq \nbfx) \mathbf{Y}_{JJ}^M(\hat{\bfx}) \cdot{\boldsymbol \calj}(\bfx)
\nonumber
\\ 
T_J^{M \, \mathrm(el)} =&  \int d^3 \bfx \, 
\Big[ 
\sqrt{J+1 \over 2 J +1} j_{J-1} \left(\nbfq \nbfx\right) \mathbf{Y}_{J \, J-1}^M(\hat{\bfx}) \cdot {\boldsymbol \calj}(\bfx) 
-\sqrt{J \over 2 J +1} 
j_{J+1}\left(\nbfq \nbfx\right) \mathbf{Y}_{J \, J+1}^M(\hat{\bfx})\cdot {\boldsymbol \calj}(\bfx)
\Big].
\label{tmatrixJ}
\end{align}
\end{widetext}

As a result, the general expression for the inelastic scattering differential cross section of a fixed target can be written in terms of the reduced transition matrix elements of the Coulomb, electric and magnetic irreducible current tensor operators

\begin{widetext}
\begin{align}
{d^2 \sigma \over d\Omega_f dE} &= { e^2 \over 2 \pi} {|\bfp_f |  \over |\bfp_i|}\,
{1\over q^4} {1 \over 2 J_\a +1} 
\Bigg[
C_L(q) \, \sum_{J=0}^{\infty} 
\left| \langle \Phi_\b ||T_J^{\mathrm(coul)} || \Phi_\a \rangle  \right|^2 
+ C_T(q) \sum_{J=1}^{\infty} 
\Big(
\left| \langle \Phi_\b || T_J^{\mathrm(el)} || \Phi_\a \rangle \right|^2  + \left| \langle \Phi_\b || T_J^{\mathrm(mag)} || \Phi_\a \rangle \right|^2
\Big) \Bigg] \, . 
\label{QEDDCSFinal}
\end{align}
\end{widetext}
This is the full multipole expansion of the differential cross section which applies any energy loss and momentum transfer. 

\subsubsection{Kinematics}
In the collision laboratory frame, the kinematics are defined by the scattering angle $\theta$ and the energy loss $E$ of the incident electron to the target. It is common to exchange the variable $\theta$ for $\nbfq$. The momentum transfer is

\begin{equation}
\nbfq^2 = (|\bfp_i |-| \bfp_f |)^2(1+\tilde{\theta}^2) \, , \,\, \tilde{\theta}^2 = \frac{4 |\bfp_i | \, |\bfp_f|}{(| \bfp_i |  - |\bfp_f |)^2} \sin^2 \frac{\theta}{2} \,,
\label{tildethetadef}
\end{equation}
where $\tilde{\theta}$ depends on the collection angle but scaled appropriately with the energy loss, $E$. For low $\theta$ and $E$ this expression reduces to the scaled angle which was defined in \cite{Schattschneider2005} for the investigation of relativistic effects in the cross section

\begin{equation}
\tilde{\theta} \simeq \frac{|\bfp_i|^2}{E_i^2} \frac{\theta}{\epsilon}
\end{equation}
where $\epsilon = \frac{E}{E_i}$. In the laboratory frame, the longitudinal and transverse coefficients can be written as

\begin{align}
C_L(q) &= {1 \over 2} \left( 1- {E^2 \over \nbfq^2} \right)^2 \left(4 E_i^2-4 E_i \, E + E^2-\nbfq^2\right) 
\nonumber \\
 &= \frac{1}{2} {\left(1 - {|\bfp_i|^2 \over (|\bfp_i| - |\bfp_f|)^2}+ \tilde{\theta}^2 \right)^2 \over (1+\tilde{\theta}^2)^2} 
 \nn\\ & \times \left[ (2 E_i-E)^2 - (|\bfp_i |-| \bfp_f |)^2 (1+\tilde{\theta}^2) \right]
 \label{cltheta}
\end{align}
\begin{align}
C_T(q) 
&={2 |\bfp_i | |\bfp_f| \, \sin^2{\theta \over 2} \over \nbfq^2} \left[ \left(|\bfp_i|+|\bfp_f|\right)^2 -2 |\bfp_i| | \bfp_f | \, \cos^2{\theta \over 2}\right]
\nn \\
 &+ \frac{(|\bfp_i |-| \bfp_f |)^2 - E^2}{2}
 \nn \\ &= {\tilde{\theta}^2 \over 2(1+\tilde{\theta}^2)} \left( \bfp_i^2 + \bfp_f^2+ {(|\bfp_i |-| \bfp_f |)^2 \over 2} \tilde{\theta}^2\right)
\nn \\
&
+ {(| \bfp_i | - |\bfp_f|)^2-E^2 \over 2}
\label{cttheta}
\end{align}
We observe that the transverse part becomes important for large scattering angles and energy loss. In case of such experiments as reflection EELS, the outgoing electrons are detected at large scattering angles, transverse excitations of the target become more accessible. In contrast, at low incident electron energy, the relativistic effects are small.


\subsection{\label{subsubsec:photoionization} Photoionization}
The ionization cross section of an atom by a high-energy photon is relevant for XAS. EELS and XAS signals contain similar information about the spectral function of the target, and it is common to be compared. The ionization cross section is calculated by considering an on-shell photon interacting with the target. The transition matrix element is

\begin{equation}
|T_{\b ; \a, k }|^2 = \frac{1}{4 \pi E} | \calj^\m \, \varepsilon_\m(k,\lambda) |^2 \,.
\end{equation}
$\varepsilon_\mu(k,\lambda)$ is the photon polarization vector and $k^0=|\boldsymbol{k}| = E_f-E_i = -E $. We can consider here the $\boldsymbol{k} \cdot \boldsymbol{\varepsilon}(k, \l)=0$ and $\varepsilon^0(k, \l)=0$. Following the same steps as in Sec. \ref{subsec:multiexp} and summing over all the photon polarizations, the multipole expansion gives the inelastic cross section of the photon-atom scattering

\begin{align}
\sigma_\g(E) &= (2 \pi)^4 {1 \over 2 J_\a +1} \sum_{M_\a \, M_\b} |T_{\b ; \a, k }|^2 =
\frac{(2 \pi)^3}{2 E} {1 \over 2 J_\a +1}
\nn\\ & \times\sum_{J=1}^{\infty} 
\left(
\left| \langle \Phi_\b || T_J^{\mathrm(el)} || \Phi_\a \rangle \right|^2+ \left| \langle \Phi_\b || T_J^{\mathrm(mag)} || \Phi_\a \rangle \right|^2
\right)
\label{photoionization}
\end{align}
where the full expansion of the $T$-matrix elements in multipoles is given by Eq.~(\ref{tmatrixJ}). The transverse parts of the two cross section expressions have the same multipole structure. We will later show that in the low-momentum transfer limit the leading term in $T^{(e)}$ is similar to the dipole approximation of $T^{(coul)}$, which explains the correspondence of  the XAS signal and EELS signals. 

The calculated multipole expansion (Sec. \ref{subsec:multiexp}) of the photon-atom interaction is of particular interest for inelastic x-ray scattering (IXS) experiments, where contributions beyond dipole are important \cite{Huotari_2015}. The multipole expansion of the IXS cross section was studied in \cite{PhysRevB.86.035138},  where the target Hamiltonian was expanded in the powers of $1/c$.


\section{\label{sec:atomicstructure} Atomic Structure}
When inner atomic shells are excited, the final state encompasses both resonances and a continuous spectrum. This study mainly focuses on transitions to the continuum, which are utilized in EELS data processing to determine the elemental composition of materials. In this case, neighboring atoms modify the near-edge fine structure but not the continuum spectrum. The detailed description of the near-edge fine structure typically requires additional effects like the crystal-field effects for ionic crystals (see Sec. \ref{sec:excDCS}) or a simulation of the lattice cell  \cite{PhysRevLett.85.1298, pantelides1, pantelides2}, depending on the EELS edge which is analyzed. We here examine the continuum component of the spectral function by considering the target of the scattering process as an atom characterized by the current operator $\calj$.  The initial state is the atomic ground state, while the final state comprises an ion with an inner-shell vacancy and an ejected electron.


\subsection{\label{c}Electromagnetic current for Dirac orbitals}

The dynamics of a neutral atom of atomic number $Z$, with $N$ electrons, is described by the Lagrangian (\ref{QEDLagrangian}), with a static external nuclear current density ${\calj^{\m}_{nucl}}(\bfx) = (\rho_{nucl}(\bfx), {\bf 0})$ assuming that the nucleus magnetic moment has a small effect. 
The ansatz for the ground-state wave function of a finite fermionic system is a Slater determinant
\begin{align}
 | \Phi_\alpha \rangle &= | \caln_1 \ldots \caln_N\rangle 
 \nn  \\ & 
 = {1 \over \sqrt{N!}}\sum_P (-1)^P |\caln_{P(1)} \rangle \otimes|\caln_{P(2)} \rangle \otimes \ldots \otimes |\caln_{P(N)} \rangle \,.
\label{slater}
\end{align}
The matrix element of the electromagnetic current operator on $N$-particle fermion states can be formulated in terms of the matrix elements of the one-particle base orbitals $| \caln_i \ra$, where $\caln_i$ denotes the quantum numbers of the $i$-th relativistic electron in a central field $\caln=\{ n, \k, m \}$. 

The ansatz for the single-particle orbitals is taken to be the central field Dirac spinors

\begin{equation}
\psi_\caln(\bfx) = 
\begin{pmatrix}
f_\caln(\bfx) +i \, g_\caln(\bfx)  \\
f_\caln(\bfx) -i \, g_\caln(\bfx)
\end{pmatrix} \,,
\label{eq:diraceigen}
\end{equation}
where $f_\caln(\bfx)=\Omega_{\k\,m}({\hat \bfx}) F_\caln(r)$, $g_\caln(\bfx)= \Omega_{-\k\,m}({\hat \bfx}) G_\caln(r)$,
\begin{equation}
 \Omega_{\k\,m}({\hat \bfx})= 
\begin{pmatrix}
C_{j \mp {1\over 2}\,, {1\over 2}} (j,m\; m-{1\over2}, {1\over2}) Y^{m-{1\over 2}}_{j\mp {1\over 2}} \\
C_{j \mp {1\over 2}\,, {1\over 2}} (j,m\; m+{1\over2}, {1\over2})
Y^{m+{1\over 2}}_{j\mp {1\over 2}}
\end{pmatrix}
\label{sphericalspinors}
\end{equation}
and $\Omega_{-\k\,m}({\hat \bfx}) = (\boldsymbol{\sigma} \cdot \bfx)\, \Omega_{\k\,m}({\hat \bfx})$. 

The current matrix element between two orbitals is defined as
\begin{align}
\calj_{\caln \calm}^\m(x) &= \langle \calm | \calj^\m(x) | \caln \rangle 
\nn \\ &= - i e \, e^{i (E_\calm-E_\caln) t} \overline{\psi}_\calm(\bfx) \gamma^\m \psi_\caln(\bfx)
\label{currentMEDirac}
\end{align}
The electromagnetic current components in terms of the wave-function ansatz in (\ref{eq:diraceigen}) are readily found:
\begin{align}
\calj_{\calm \caln}^0(\bfx) &=-e \, \left( f_\calm^\dagger(\bfx) f_\caln(\bfx) + g_\calm^\dagger(\bfx) g_\caln(\bfx) \right)
\label{jdirac0} \\
{\boldsymbol \calj}_{\calm\caln}(\bfx) & = i e\, \left( f_\calm^\dagger(\bfx) {\boldsymbol \sigma} g_\caln(\bfx) + g_\calm^\dagger(\bfx) {\boldsymbol \s} f_\caln(\bfx) \right) 
\label{jdiraci}
\end{align}
Equations~(\ref{tmatrixJ}) lead to the $T$-matrix elements in terms of radial Dirac wave functions. The calculation of the irreducible matrix elements of the current operator with the vector spherical spinors is shown in detail in the appendix \ref{app:multipoleexpansion}. The final result reads as

\begin{widetext}
\begin{align}
\langle \calm ||T_J^{\mathrm(coul)} || \caln \rangle &= 
-e \langle \k_\calm || Y_J (\hat{x}) || \k_\caln \rangle
\int dr \, r^2 
j_J\left(\nbfq r \right) \left[ F_\calm^\dagger(r) F_\caln (r) + G_\calm^\dagger (r) G_\caln(r)
\right] 
\nonumber \\ 
\langle \calm || T_{J}^{(mag)}  || \caln \rangle &=-i \, e {(\k_\caln+\k_\calm) \over \sqrt{J(J+1)}}
\langle \k_\calm || Y_{J} || -\k_\caln \rangle
\int dr \, r^2\, j_J\left(\nbfq r\right) 
\left( F_\calm^\dagger(r) G_\caln(r)+G_\calm^\dagger(r) F_\caln (r) \right)
\label{Tcoulme} \\ 
\langle \calm || T_{J}^{(el)}  || \caln \rangle
&=-i\, e \, \sqrt{J + 1 \over J}
\langle \k_\calm \vert \vert Y_{J}^{M}\vert \vert \k_\caln \rangle
\int dr \, r^2 \, 
\left[
{\k_\caln - \k_\calm \over J+1} 
\left( {j_J\left(\nbfq r\right) \over \nbfq \,r } + j_J^\prime(\nbfq \, r) 
\right)
\right.
\nonumber \\ 
& \left. \times \left(F_\calm^\dagger(r) G_\caln(r)+G_\calm^\dagger(r) F_\caln(r) \right)
+ J \, {j_J\left(\nbfq r\right) \over \nbfq \,r } 
\left(F_\calm^\dagger(r) G_\caln(r)-G_\calm^\dagger(r) F_\caln(r) \right)
\right]
\nonumber
\end{align}
\end{widetext}
where the reduced matrix element of the spherical harmonic $Y_J$ is defined as 
\begin{equation}
\langle \k_\calm || Y_{J} || \k_\caln \rangle =
{(-1)^{j_\calm-{1\over2}} \over \sqrt{4 \pi}} \, \left[ j_\calm, \, J, \, j_\caln \right]^{1\over 2} 
\begin{pmatrix}
j_\calm & J & j_\caln \\
{1\over 2} & 0 & -{1\over2}
\end{pmatrix}
\label{Ymatrixelement}
\end{equation}
These are generic expressions for the $T$-matrix, describing the scattering of a relativistic electron from a fully relativistic target, formulated in terms of Dirac angular-momentum eigenstates and one-dimensional integrals of the radial orbitals. These expressions are easy to use for the calculation of core-loss EELS spectral shapes with different models for the target dynamics. We observe that the dominant contributions of the small spinor components primarily arise from the mixing terms in the transverse matrix elements.

Equations~(\ref{Tcoulme}) refer to single-particle states, which are utilized to compute matrix elements for many-body states expanded in a one-particle orbital basis, as shown in appendix \ref{app:matrixelement}. This formalism is directly applicable to configuration interaction states, which are represented as linear combinations of Slater determinants  \cite{JONSSON2007597}. The expressions for $T^{(e)}$ and $T^{(m)}$ also appear in the multipole expansion of the relativistic photoionization cross section and the x-ray emission rate as it follows from, Eq. (\ref{photoionization}),  which has been studied in \cite{doi:10.1080/00018737000101191, IPGrant_1974, merstorf2023nonlorentzianatomicnaturallineshape}.

\subsection{\label{subsec:lowq} Low-momentum transfer limit}
The relevant energy scales in an EELS experiment are controlled by the beam energy, momentum transfer, and energy loss. Typically, low-momentum transfer and low-energy loss experiments are performed in a TEM with a relativistic beam. Our generic result in Eqs.(\ref{QEDDCSFinal}), applies to arbitrarily high-energy loss and momentum transfer.  In this section, we derive the long-wavelength limit of the full cross-section expression by expanding Eqs. (\ref{QEDDCSFinal}) and (\ref{Tcoulme}) in terms of the energy loss and scattering angle. For nonrelativistic atoms, we use the Pauli approximation for the orbital wave functions:
\begin{equation}
G(r) \simeq -\frac{1}{2 m } \left( {d F(r) \over dr} + {\kappa+1 \over r} F(r) \right) \,.
\label{eq:pauliwf}
\end{equation}
The expansion of the Coulomb matrix element $T^{(coul)}$ is straightforward, since the second term is second order in $G(r)$. Consequently, the small spinor component significantly impacts the electron-target interaction when transverse effects are included. Using Eq.~(\ref{eq:pauliwf}), the magnetic part of the $T$ matrix becomes
\begin{widetext}
\begin{align}
\langle \calm || T_{J}^{(mag)}  || \caln \rangle =& \frac{i \, e}{ 2 m} {(\k_\caln+\k_\calm) \over \sqrt{J(J+1)}}
\langle \k_\calm || Y_{J} || -\k_\caln \rangle \,
\int_0^{\infty} dr \left( (\k_a+\k_b) \,r \, j_J(\nbfq r) - r^2{d \over dr} j_J(\nbfq r) \right) F_\calm(r) F_\caln(r) 
\end{align}
\end{widetext}
The ionized electron wave function is nonrelativistic when the momentum transfer is small, $\nbfq r \ll 1$, where $r$ is the characteristic radial distance of the continuum wave-function overlap with the core-hole shell. In this case, the spherical Bessel function can be expanded to
\begin{equation}
j_J(\nbfq r) = {(\nbfq r)^J \over (2 J+1)!!} \,.
\label{besselExpansion}
\end{equation}
With some algebra, the long-wavelength approximation of the magnetic part reduces to
\begin{widetext}    
\begin{align}
\langle \calm || T_{J}^{(mag)}  || \caln \rangle = &
{i \, e \over 2 m} {(\k_\caln+\k_\calm) \left( \k_\caln+\k_\calm - J \right)  \over (2J+1)!! \sqrt{J(J+1)}}
\langle \k_\calm, || Y_{J} || -\k_\caln \rangle \,
| \bfq |^J
\times \int dr \, r^{J+1} F_\calm(r) F_\caln(r) \,.
\end{align}
The electric part leads to
\begin{align}
\langle \calm || T_{J}^{(el)}  || \caln \rangle 
=&{i\, e \over 2 m}\, \sqrt{J + 1 \over J}
\langle \k_\calm \vert \vert Y_{J} \vert \vert \k_\caln \rangle \,
{2 J \over (2J +1)!!} | \bfq |^{J-1}
\nonumber \\ 
&
\times \int_0^{\infty} dr \, r^{J+1 \over 2 } F_\calm(r)
\left({d \over dr} + {(\k_\caln - \k_\calm)(\k_\caln+\k_\calm+1) \over 2 J \, r} \right) \left(r^{J+1 \over 2 }F_\caln(r) \right)
\label{TelectricExpansion}
\end{align}
\end{widetext}
This is an extension of the standard electric dipole ($J=1$) transition matrix element in the velocity gauge \cite{IPGrant_1974}. The total derivative terms vanish at the boundaries of integration since $F(r) \to r^{\ell+1}$ for $r\to 0$. The exception is the monopole transition, which is not relevant in the current context because it is canceled by the nucleus contribution in the cross section.  If the Hamiltonian is local, this can be rewritten in the length gauge. However, in the case of a nonlocal potential, such as the exchange potential, the situation becomes more complicated. Higher-order many-body effects restore the gauge invariance of the transition matrix elements in multiparticle systems \cite{Amusia1971}. Equation~(\ref{TelectricExpansion}) leads to the relation of the electric $T$ matrix to the low-momentum Coulomb $T$ matrix.
\begin{align}
\langle \calm || T_{J}^{(el)}  || \caln \rangle  \simeq
{E \over \nbfq}\sqrt{J+1 \over J}
\langle \calm || T_{J}^{\mathrm(coul)}  || \caln \rangle \,.
\label{telexp}
\end{align}
This relation has been implicitly applied in the EELS literature as the relativistic cross section. Typically, the long-wavelength approximation is done already at the operator level \cite{PhysRev.102.385}. The derivation above shows in detail how the long-wavelength approximation is derived from the full multipole expansion.

The expansion of Eqs.~(\ref{cltheta}) and (\ref{cttheta}) for low $E/E_i$ and $\theta$ together with Eq.~(\ref{telexp}) leads to the low-energy limit of the full $T$ matrix: 
\begin{widetext}
\begin{align}
|T_{p_f,\,\beta \,;\, p_i,\,\alpha}|^2  \simeq
 & \left( C_L(q) + \frac{E^2}{\nbfq^2} \frac{J+1}{J} C_T(q) \right) \,
\left| \langle \Phi_\b ||T_{J=1}^{\mathrm(coul)} || \Phi_\a \rangle  \right|^2
\nonumber \\
\simeq 
& \Bigg[
 2 {m^4 \over E_i^2}\left(1+\frac{E}{E_i} 
 +{2 E_i^4 + 2 E_i^2 m^2 - 3 m^4 \over 4 m^2 p_i^2} {E^2 \over E_i^2} + \ldots
 \right)
\nonumber \\
& +  2 \left( {p_i^2(E_i^2+m^2) \over E_i^2} + \frac{E_i^4+m^4}{E_i^2}\frac{E}{E_i} +
{2 E_i^6 - 6 E_i^4 m^2 -2 E_i^2 m^4 + 3 m^6 \over 4 E_i^2 p_i^2} {E^2 \over E_i^2} + \ldots\right)
\tilde{\theta}^2 
\nonumber \\
&+ \mathcal{O}\left( \tilde{\theta}^4\right)
\Bigg]\,
\left| \langle \Phi_\b ||T_{J=1}^{\mathrm(coul)} || \Phi_\a \rangle  \right|^2
\label{dipoleexpansionT}
\end{align}
\end{widetext}
We emphasize that this is a good approximation for small scattering angles and small energy losses \cite{berestetskii1982quantum}. At larger values of the energy loss and scattering angle, the higher-order terms in the above expansion become important.

Core-loss EELS experiments are typically performed at larger collection angles, especially for lower-resolution measurements aiming at elemental quantification. The acquisition of EELS data at higher energy loss has become more accessible after recent developments in TEM technology \cite{PhysRevApplied.23.054095}.
Such experiments typically target the analysis of the extended fine structure of $K$ edges to extract information such as the coordination number and nearest-neighbor distances. In this context, the full cross section given in Eqs.~(\ref{QEDDCSFinal}) and (\ref{Tcoulme}) is relevant for the removal of the continuous part of the differential cross section and the extraction of the interference pattern, due to the scattering of the ionized electron from the neighboring atoms \cite{10.2138/rmg.2014.78.2}.

\subsection{\label{subsec:atom}Dirac-Hartree-Fock}
Multielectron atoms are studied systematically within the framework of many-body theory and Green's functions \cite{fetter2003quantum, PhysRevA.74.062503}. 
The wave-function ansatz for a finite system of fermions is given by Eq. (\ref{slater}). The equations of motion of the one-particle orbitals in coordinate space follow from Dyson's equation
\begin{align}
\Big( & -i {\boldsymbol  \alpha} \cdot {\boldsymbol  \nabla} +  \beta \, m  + V_{nucl}(\bfx) \Big) \psi_\caln(\bfx)
\nn \\ &
+\sum_{\caln'} \,\Sigma(\caln, \caln' ; \varepsilon_\caln )
\psi_{\caln'}(\bfx) 
= \varepsilon_\caln \psi_\caln(\bfx) \,,
\label{cordeom}
\end{align}
where $\boldsymbol{\alpha} = -\gamma^0 {\boldsymbol \gamma}$, $\beta = i \gamma^0$, and $\Sigma$ is the irreducible self-energy. The self-energy includes the direct and exchange contributions from the electron-electron interaction and higher-order terms in the vertex corrections. In DHF, the vertex function is taken to be zero, meaning that no vertex corrections are included to the self-energy beyond the bare interaction
\begin{align}
\Sigma^{DHF}(\caln, \caln') 
& =
\sum_{\calm < F}\la \caln \, \calm | V_{ee} | \caln' \, \calm \ra \,,
\label{selfEnergy}
\end{align}
where the antisymmetrized electron-electron matrix element is defined in Eq.~(\ref{eepotential}). $V_{ee}(|\bfx_i - \bfx_j|)$ is the electron-electron interaction. At the lowest order, this is the Coulomb potential. For inner-shell electrons of heavier atoms the Breit interaction becomes non-negligible:

\begin{equation}
V_{ee}^{(Breit)}(|\bfx_i- \bfx_j|) \simeq -\frac{
{\boldsymbol \alpha}_i \cdot {\boldsymbol \alpha}_j + {\boldsymbol \alpha}_i \cdot \boldsymbol{\hat{\bfx}_{ij}} \,
{\boldsymbol \alpha}_j \cdot \boldsymbol{\hat{\bfx}_{ij}} }{|\bfx_{ij}|}  \,.
\label{eq:breit}
\end{equation}
The interaction at the one-loop level will not play a leading role in our results. The form of the effective one-loop potentials is discussed in \cite{PhysRevA.72.052115}. $\mathrm{V_{nucl}}$ is the nuclear potential derived from a Fermi distribution for the nuclear charge density \cite{BRACK1985275},

\begin{equation}
\rho(r) = \frac{\rho_0}{1 + e^{B(r - r_0)}} \, ,
\nonumber
\end{equation}
with $B = \frac{4 log 3}{T}$, $T$ is a nuclear thickness, taken to be a constant equal $2.3$ fm. $r_0$ is defined in terms of atomic mass as
\begin{equation}
r_0 = 0.836\, A^{1/3} + 0.57\,.
\nonumber
\end{equation}
$\rho_0$ is determined by the total nuclear charge 
\begin{equation}
Z\, e = \rho_0 \int_0^{\infty}{1 \over 1 + e^{B(r - r_0)}} \,.
\nonumber
\end{equation}


\subsection{\label{sec:finalstate} Final state and the core hole}

In the case of inner-shell ionization, the main collective effect is the atomic core rearrangement in the final state of the atom, driven by the generated vacancy in a shell localized quite deep in energy \cite{PhysRevB.25.5150, Rehr_2005, Amusia1981, refId0, ELindroth_1993, RevModPhys.75.35, Derevianko_2004, PhysRevB.17.560, Zarkadas}. An intuitive physical argument motivates this picture. An ionized electron leaving the atom moves in the field of the created ion where the other bound electrons rearrange due to the presence of the core hole. The effect is particularly important if the ionized electron does not leave the atom well before the bound electrons have the time to relax due to the vacancy created in a shell $\call$. An order of magnitude estimation shows that for ionized electrons with small kinetic energy, i.e. close to the ionization threshold, this effect plays a role. The relaxation time can be estimated by the energy difference of the neutral atom ground-state energy $E_{\mathrm{atom}}$, the relaxed ion energy $E_{\mathrm{ion}}$ and the absolute value of the vacancy Hartree-Fock eigen-energy $E_\call$:
\begin{equation}
\Delta \tau_i \sim {1 \over E_{atom}-E_{ion}-E_\call} 
\label{eq:relaxationtime}
\end{equation}
Meanwhile, the characteristic time that a slowly moving ionized electron spends close to the atomic cloud is
\begin{equation}
t_{\epsilon} \sim a_B \sqrt{ m_e \over 2 \epsilon} \,.
\end{equation}
For $t_{\epsilon} \gtrsim  \Delta \tau_i$ or, equivalently, for low energy of the ionized electron, the relaxation of the atom due to the vacancy is important. This effect is smaller for fast-moving ionized electrons, and the state of the relaxed system approximates the frozen core solution. This is natural since the electromagnetic field of the relaxed and frozen ion has the same asymptotics at large distances.

The dynamics of a particle hole moving in a multiparticle medium is typically studied within RPAE which reproduces the expected pole structure of the polarization propagator and has been proven useful in the study of atomic excitations \cite{Amusia1971, WRJohnson_1980, Amusia1975}. In appendix \ref{app:rpa}, we show that an equivalent approach for inner-shell ionization is to find self-consistent solutions for the ion state with a core hole, the so-called final-state rule or delta self-consistent field method. One can then solve the equation of motion of the ionized electron in the field of the fully relaxed ion,
\begin{align}
    &{\widetilde H}_{DHF}\, \widetilde{\psi}_\caln(\bfx) -\int d\bfx' \,
    {\widetilde \psi}^{\dagger}_{\call}(\bfx') V_{ee}(| \bfx-\bfx'|) {\widetilde \psi}_{[\call} (\bfx') \, {\widetilde \psi}_{\caln]} (\bfx)
   \nn \\
   &= \widetilde{\epsilon}_{\caln} \widetilde{\psi}_\caln(\bfx)
\label{dhfcorehole}
\end{align}
where the vacancy is at the orbital $| \call \rangle$, and $\widetilde{H}_{\mathrm{DHF}}$ denotes the Hamiltonian in terms of the relaxed orbitals with the kinetic and interaction terms which are mentioned in Sec. \ref{subsec:atom}. The bound and continuum wave functions calculated in the presence of the core-hole potential must be projected into the orthogonal subspace of the initial neutral atom ground state, see (Appendix~\ref{app:corehole}). This is necessary to obtain a physically meaningful transition matrix element, one which yields no transition for a trivial operator \cite{Gell-Mann:1953dcn}

The many-body effects on the calculation of the transition matrix element of an excitation operator due to the core-hole potential are manifest in Eq.~(\ref{MBMatrixElement}). The core-hole interaction with the atomic orbitals adds corrections to the matrix element which are shown in the perturbative diagrammatic expansion of Fig.~\ref{fig:MBRPA}. Those include the propagation and interaction of the electron-hole pair within the polarized medium. The quantum numbers of an inner-shell vacancy are not modified since its mixing with other shells is suppressed by their large energy separation and the weak spatial orbital overlap. This is quantitatively shown within RPAE in appendix~\ref{app:rpa}. The backward-propagating terms of full RPAE are estimated to be small, Eq. (\ref{tdalimit}), and the RPAE equations of motion therefore decouple and reduce to the relaxed DHF. A quantitative comparison of the virtual intershell and intrashell  electron-hole pair contributions to the ionized electron-hole pair demonstrates the suppression of such terms. The self-energy of the vacancy also contributes to the transition matrix element, as seen in the diagram of Fig.~\ref{fig:MBPol}. These two contributions summarize the two main correlation effects in inner-shell ionization.

\section{\label{sec:results} Numerical Results}

\subsection{Numerical methods}
The DHF system of coupled differential equations is solved numerically in \texttt{AMBIT} \cite{ambit} for a number of atoms.  The eigenvalue problem is solved by the shooting method. A combination of exponential and linear grids is applied to the radial direction for the solution of the ordinary differential equations (ODEs). The ODEs are solved using the Adams-Moulton integration method until self-consistency is achieved. The ionized electron wave function is numerically determined in a similar fashion but it does not back-react to the bound-state orbitals of the ionized atom. The equation of motion is solved numerically with a boundary condition at $r=\infty$, which corresponds to a free outgoing spherical wave.

In the calculation of the ionization differential cross section, the occupation number of the open shells is taken to be an average over the quantum numbers of the open shell, so that spherical symmetry is kept. This approximation is not expected to have a major impact on the overall shape of the ionization differential cross section of core shells since they are expected to decouple from the valence shells \cite{indelicato_lindroth_1992}. However, vector coupling of the angular momentum is needed for the discrete excitation part of the spectra, as calculated in Sec. \ref{sec:excDCS}. 

An auxiliary central potential which resembles the isotropic part of the electron-electron interaction is commonly used as the starting point of the iterative solution of the DHF equations of motion. The noncentral part of the $V_{ee}$ potential is added during the iterative solution. The DHF auxiliary central field is chosen to be the Thomas-Fermi potential. The final result does not depend on this choice, but the convergence speed of numerical integration of the ODEs can be substantially improved by a good choice of the auxiliary potential.
\begin{figure}[!htbp] 
\begin{subfigure}[t]{0.8\linewidth}
\caption{}
\includegraphics[width=\linewidth]{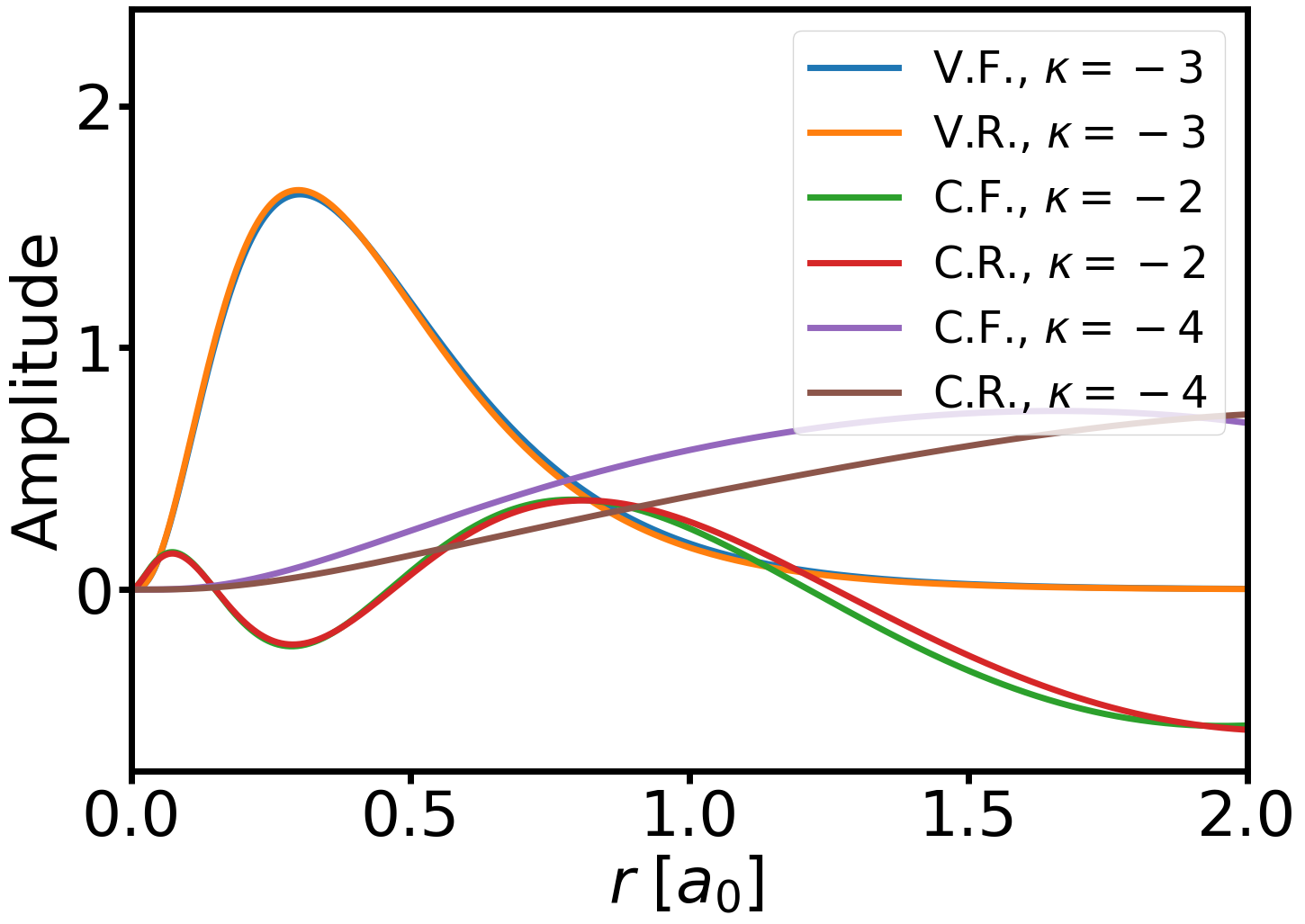}
\end{subfigure}\hspace{10pt}%
\begin{subfigure}[t]{0.8\linewidth}
\caption{}
\includegraphics[width=\linewidth]{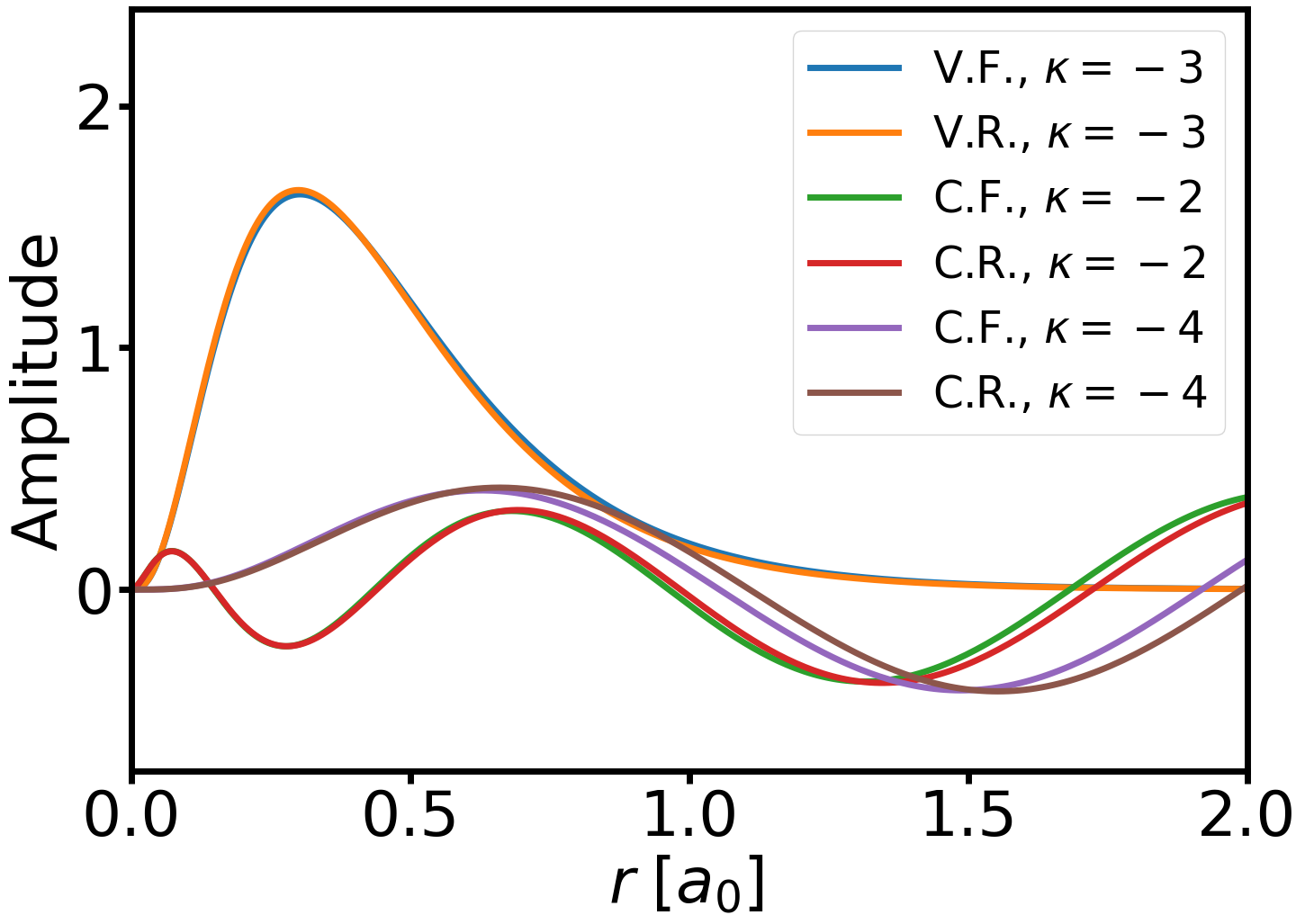}
\end{subfigure}%
\caption{The large component of the Dirac wave functions of the Ru-$3d_{5/2}$ orbital and the ejected electron in terms of the radial distance $r$ at energy 30eV (a) and 200eV (b) above the ionization threshold for the relaxed and frozen DHF approximations. The wave functions of the vacancy orbital in the frozen and relaxed cases are denoted as VF:  vacancy frozen and VR: vacancy relaxed. The continuum wave functions for $\kappa =-2,-4$ in the frozen and relaxed cases are denoted CF: continuum frozen, CR: continuum relaxed. }
\label{Ruwavefunctions}
\end{figure}
The rearrangement of the atomic orbitals in the final target state is explicitly considered by solving Eq.~(\ref{dhfcorehole}) for the final state of the bound atomic and ejected electrons. The orthogonality property of the final-state solutions is addressed in a standard perturbation theory fashion. The state that is perturbed by the core-hole interaction is projected into the initial-state subspace, as described in Appendix \ref{app:corehole}. The orbital wave functions are numerically stable and they are orthonormal to numerical precision of $\sim O(10^{-9})$. In Fig.~\ref{Ruwavefunctions}, we show the large component of the Dirac wave function of the $3d_{5/2}$, ($\kappa = -3$) Ru subshell, computed in the relaxed and frozen approximations. In the frozen approach, the final-state orbital wave functions are the same as in the initial state. Moreover, the continuum wave function of the ejected electron with $\kappa = -2, -4$ is displayed. This is the continuum state which can be reached by the leading dipole transition from $3d_{5 / 2}$. As expected, relaxed wave functions result in higher screening of the nuclear charge. The continuum wave-function amplitude grows slower and reaches its maximum farther from the nucleus in the relaxed case. In contrast, the bound-electron wave functions are pulled towards the nucleus, albeit this effect is less significant. As a result, the effective overlap between the continuum electron wave function and the corresponding ionic one is reduced, which leads to a smoother cross-section curve near the ionization energy.

\subsection{Ionization differential cross section}

\subsubsection{\label{subsubsec:FSE} Final-state effects}

We now present numerical results for the differential ionization cross sections of atoms by fast electrons, which are directly relevant to the interpretation of electron energy loss spectroscopy (EELS) data for elemental quantification in materials. Building on the preceding theoretical analysis of many-body effects in the target’s final state, we focus here on a quantitative comparison of different final-state approximations for selected elements. In the numerical calculations, the fully relativistic expression for the ionization cross section is applied, including the Coulomb, electric, and magnetic channels as shown in Eqs.~(\ref{QEDDCSFinal}) and (\ref{Tcoulme}). The infinite sum over $J$ is practically truncated at the value where the next term contributes less than $0.01\%$ to the total. The maximum value which is used in our computations is $J=4$.

\begin{figure}[!htbp]
  \begin{subfigure}[t]{0.8\linewidth}
    \caption{}
    \centering
    \includegraphics[width=\linewidth]{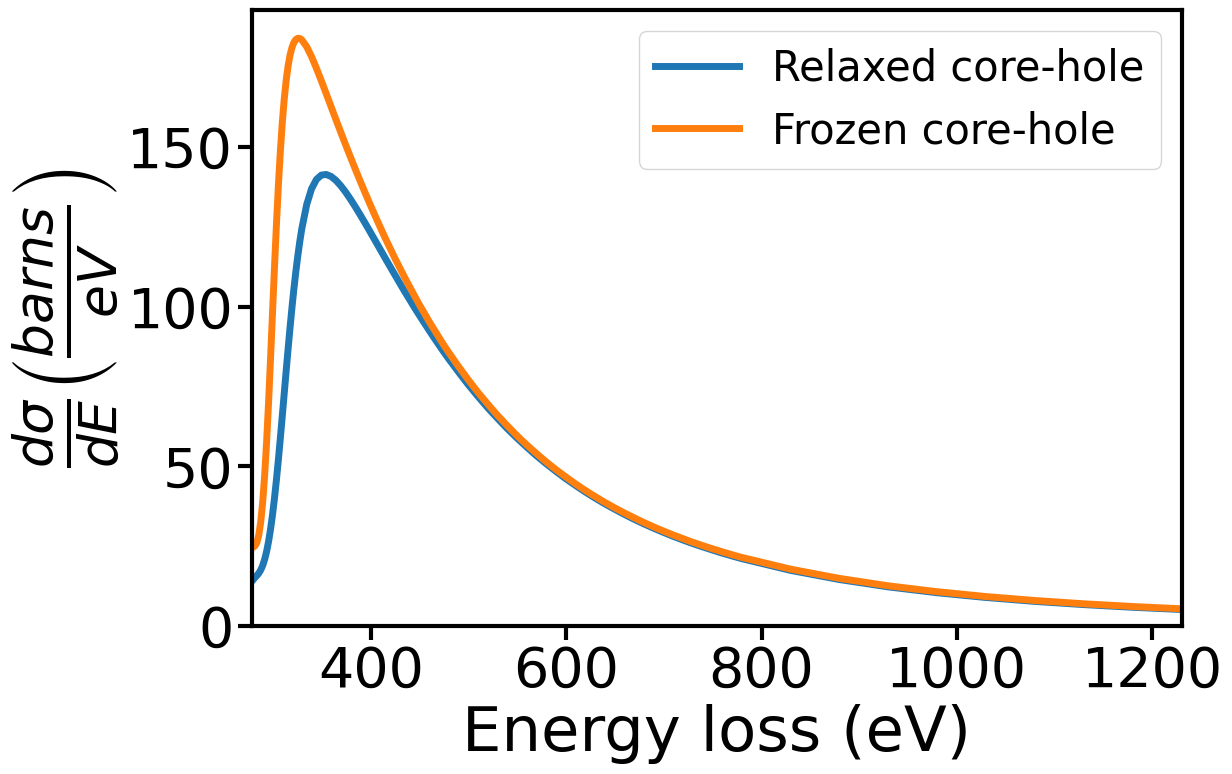}
    \label{fig:FrozenRelaxedRu:a}
  \end{subfigure}
  \vspace{10pt}%
  \begin{subfigure}[t]{0.8\linewidth}
    \caption{}
    \centering
    \includegraphics[width=\linewidth]{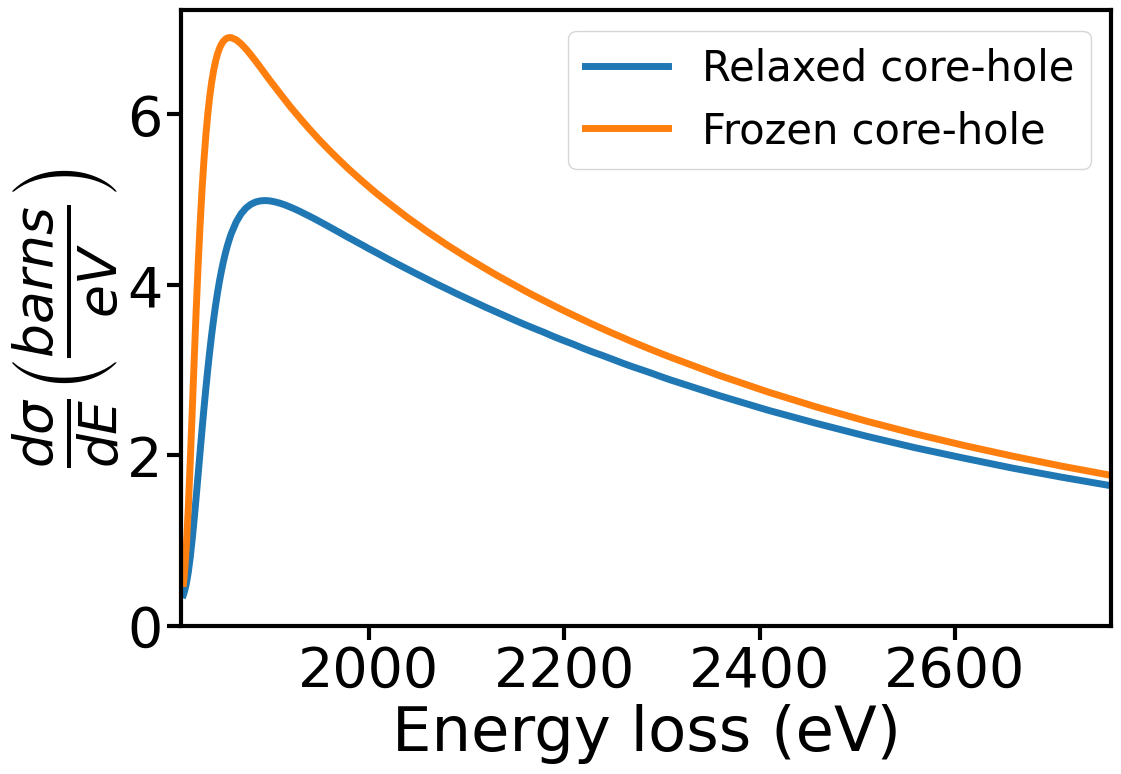}
    \label{fig:FrozenRelaxedW:b}
  \end{subfigure}
\caption{Ru-$M_5$ (a) and W-$M_5$ (b) differential cross section in terms of the energy loss of the incoming electron Eq. (\ref{QEDDCSFinal}). For each atom, the ground state is calculated within the DHF framework nd the final state is calculated in the relaxed (orange curves) and frozen (blue curves) approximations. Correlation effects such as the orbital relaxation due to the core hole make the spectrum shape smoother close to the ionization threshold compared to the frozen-core approximation. The beam accelaration voltage is $300$ keV and the collection angle $100$ mrad.}
\label{fig:FrozenRelaxedRuM}
\end{figure}

Figure~\ref{fig:FrozenRelaxedRuM} displays the differential cross sections for the $M_5$ subshell of Ru and W, respectively, calculated using two models: a fully relaxed atomic core (relaxed core hole) and a frozen core (frozen core hole). In the relaxed case, the atomic core is recalculated in the presence of the core hole, and the wave function of the ionized electron is obtained using self-consistently relaxed bound-state orbitals. In the frozen-core approximation, by contrast, the final state is described using the non-relaxed ground-state orbitals of the neutral atom, with the vacancy introduced by simply removing the ionized electron, without accounting for any relaxation due to ionization. The ejected electron wave function is calculated including both the direct and exchange interactions with the atomic electrons.  At high energies, both models yield similar asymptotic behavior, as the high-energy outgoing spherical wave experiences approximately the same potential from the atomic charge density in either case. We observe that the screening of the core hole from the atomic electrons results in the reduction of the ionization cross section near the ionization threshold.

\begin{figure}[!htbp]
  \begin{subfigure}[t]{0.8\linewidth}
    \caption{}
    \centering
    \includegraphics[width=\linewidth]{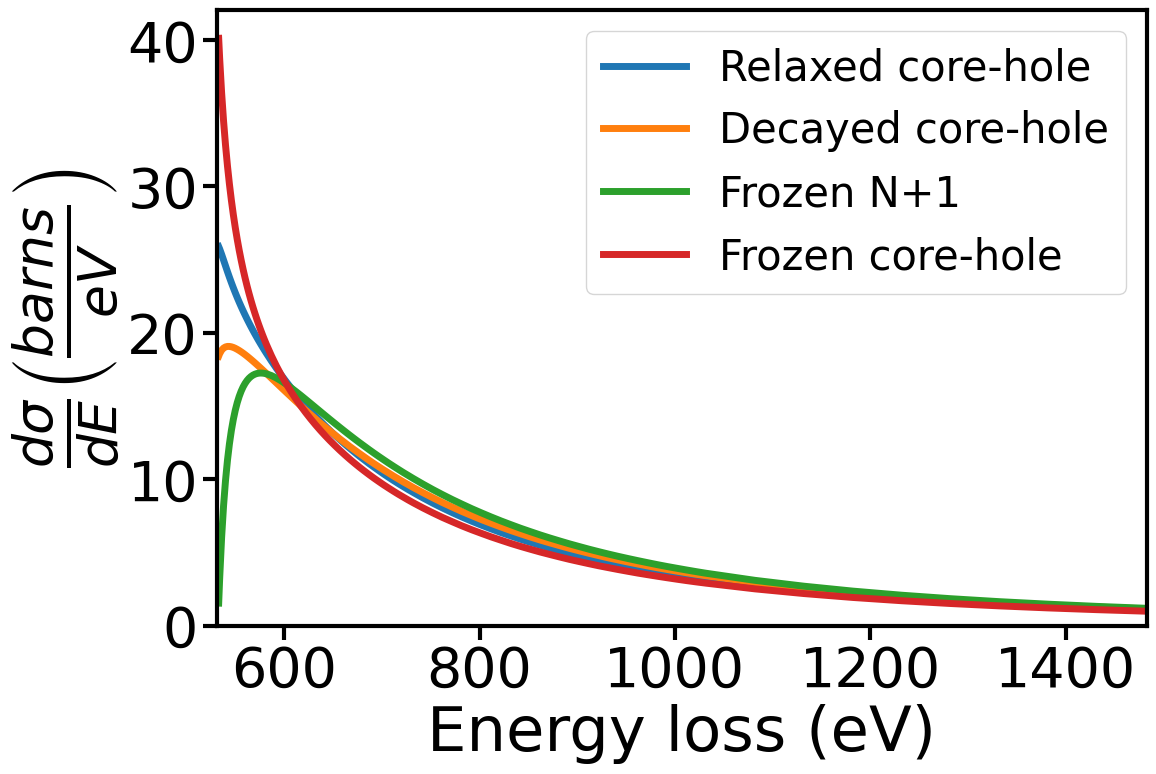}
    \label{fig:FrozenRelaxedO:a}
  \end{subfigure}
  \vspace{10pt}%
  \begin{subfigure}[t]{0.8\linewidth}
    \caption{}
    \centering
    \includegraphics[width=\linewidth]{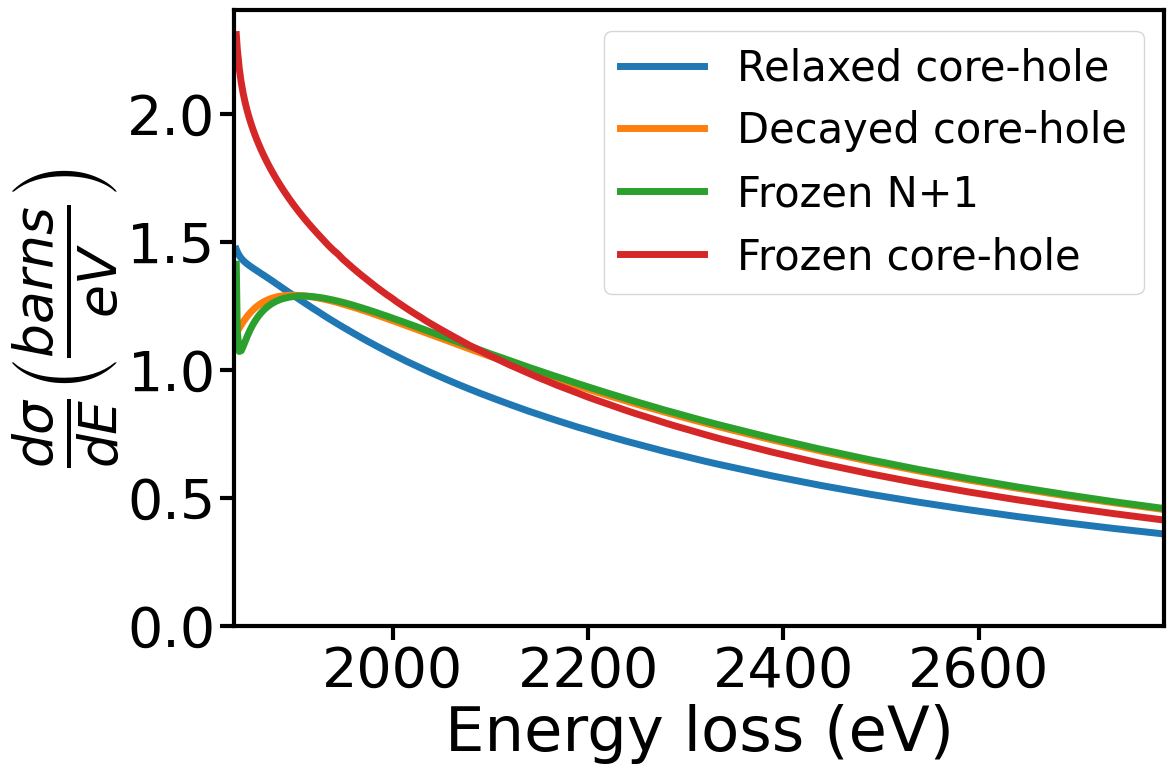}
    \label{fig:FrozenRelaxedSi:b}
  \end{subfigure}
  \caption{The comparison of the differential ionization cross sections for O-$K$ (a) and Si-$K$ (b). {\it Relaxed core-hole:} the atomic orbitals relax under the influence of the core-hole potential. {\it Decayed core hole} the core hole decays to be filled in by a valence electron. {\it Frozen $N+1$:} the final atomic bound states are the same as in the neutral atom. {\it Frozen core hole} the final state has one vacancy in the ionized shell, but all the orbitals are the same as in a neutral atom. The beam acceleration voltage is $V_\mathrm{beam}=300$ kV and the maximum scattering angle is $\theta = 100$ mrad.}
  \label{fig:FrozenRelaxedOSi}
\end{figure}

Figure~\ref{fig:FrozenRelaxedOSi} also includes results for a final-state configuration in which the core vacancy has been filled by a valence electron (decayed core hole). The final state is obtained by solving the self-consistent equations of motion for a relaxed atom with $N-1$ electrons and a vacancy in the valence shell. Although this approach allows for an approximate treatment of the core-hole decay, a fully consistent description of the final state and the post-collision interaction would also include the Auger and x-ray emission processes \cite{Amusia1981}. The decayed core-hole approximation can be relevant for deep shells of heavy elements, where the core-hole lifetime is significantly shorter than the characteristic atomic relaxation time [see Eq.~(\ref{eq:relaxationtime})]. In such cases, strong screening effects in the final state significantly influence the shape of the differential cross section, especially near the ionization threshold.

In the frozen $N+1$ case, the final-state electron configuration of the target atom is taken to be the same as that of the ground state, and the ionized electron wave function is calculated in the corresponding effective potential. In this approach, the number of electrons is not conserved during ionization, and the resulting excitation spectrum may not accurately reflect physical reality \cite{Amusia1969}. A similar approximation has been widely used in the calculation of ionization cross sections for EELS, particularly within the Hartree–Slater model, where the ionized electron is treated in the ground-state potential, which includes an empirical local exchange term \cite{PhysRevA.6.1013, Rez}.

\begin{figure}[!htbp]
\begin{subfigure}[t]{0.85\linewidth}
\caption{}
\centering{}
\includegraphics[width=\linewidth]{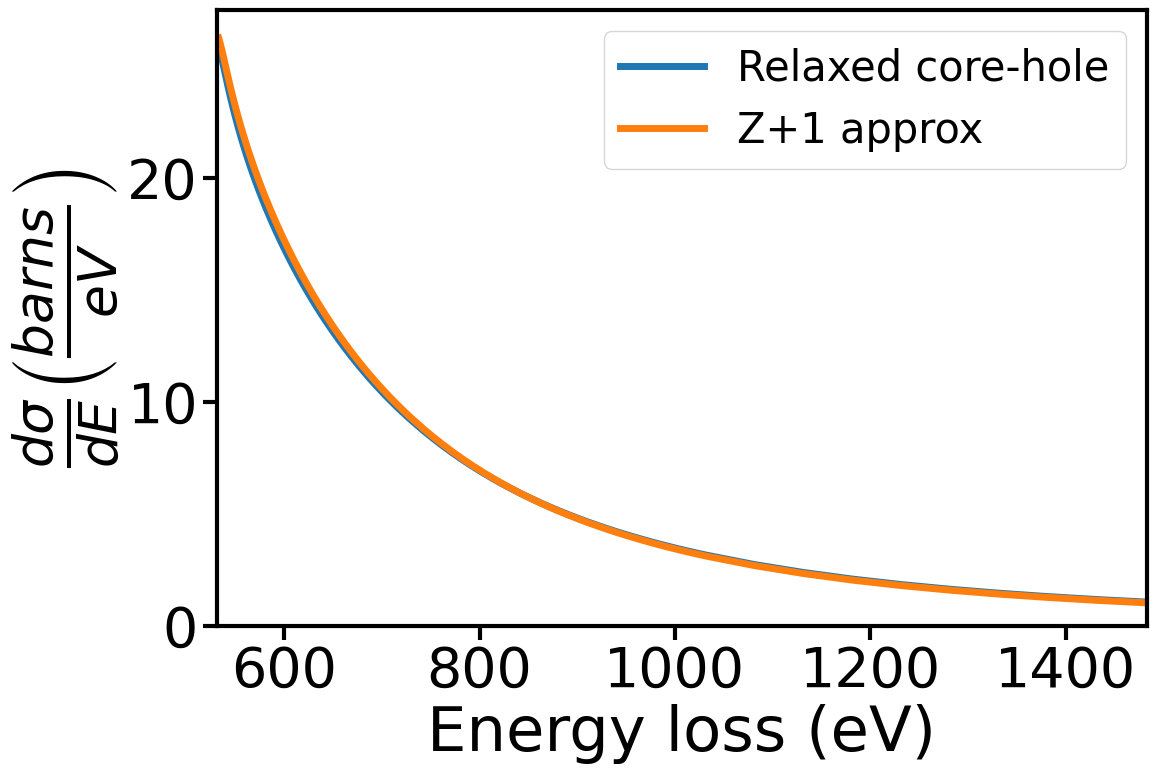}
\end{subfigure}\hspace{10pt}%
\begin{subfigure}[t]{0.8\linewidth}
\caption{}
\centering{}
\includegraphics[width=\linewidth]{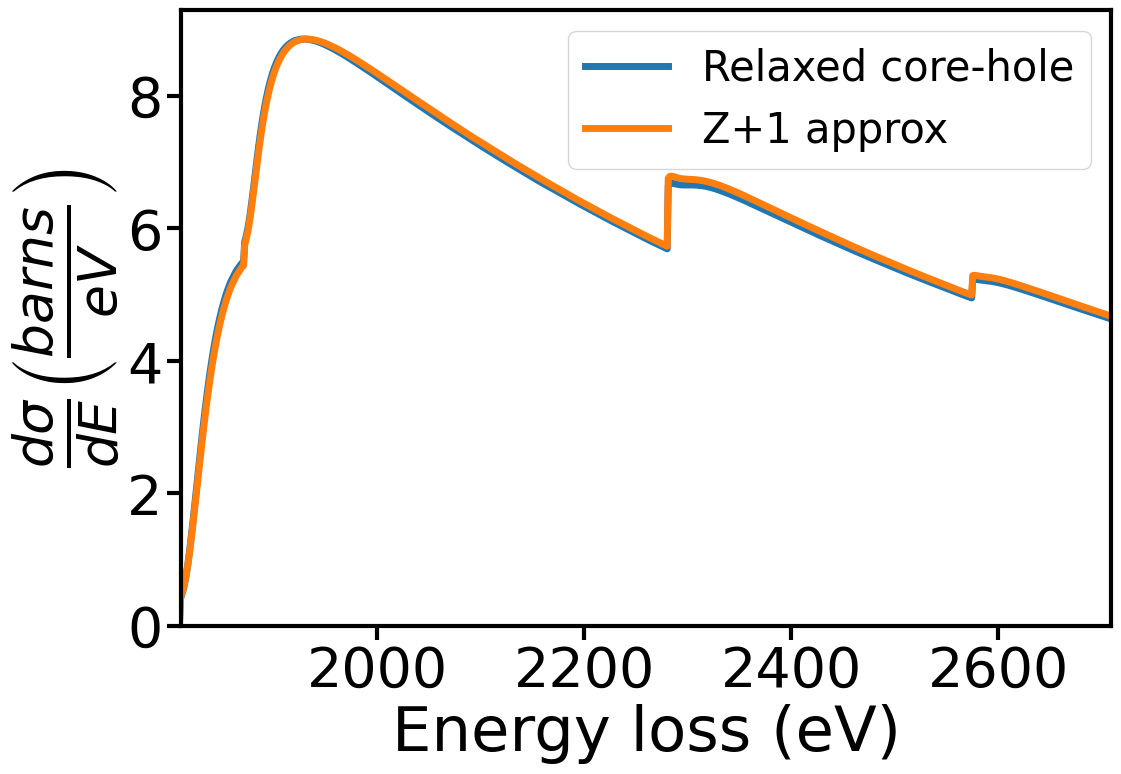}
\end{subfigure}
\caption{O-$K$ (a) and W-$M$ (b) ionization differential cross sections given by Eq. (\ref{QEDDCSFinal}) in the relaxed core-hole and $Z+1$ approximations within DHF. The beam voltage is $300$ kV and the collection angle $100$ mrad.}
\label{fig:z+1}
\end{figure}

Another method of approximating the relaxation of the atomic core in the presence of a core hole is the so-called $Z+1$ approximation \cite{pantelides1}. In this approach, the final atomic state is calculated self-consistently using a nuclear charge of $Z+1$ and  $Z$ atomic electrons with the same quantum numbers as in the neutral atom. The increased nuclear charge induces a redistribution of the electron density, effectively mimicking the screening effects present in the relaxed core-hole configuration. As shown in Fig.~\ref{fig:z+1}, this approximation yields results that closely agree with those obtained using the fully relaxed core-hole method.

\subsubsection{Relativistic effects}
\begin{figure}[!htbp] 
\begin{subfigure}[t]{0.9\linewidth}
\caption{}
\includegraphics[width=\linewidth]{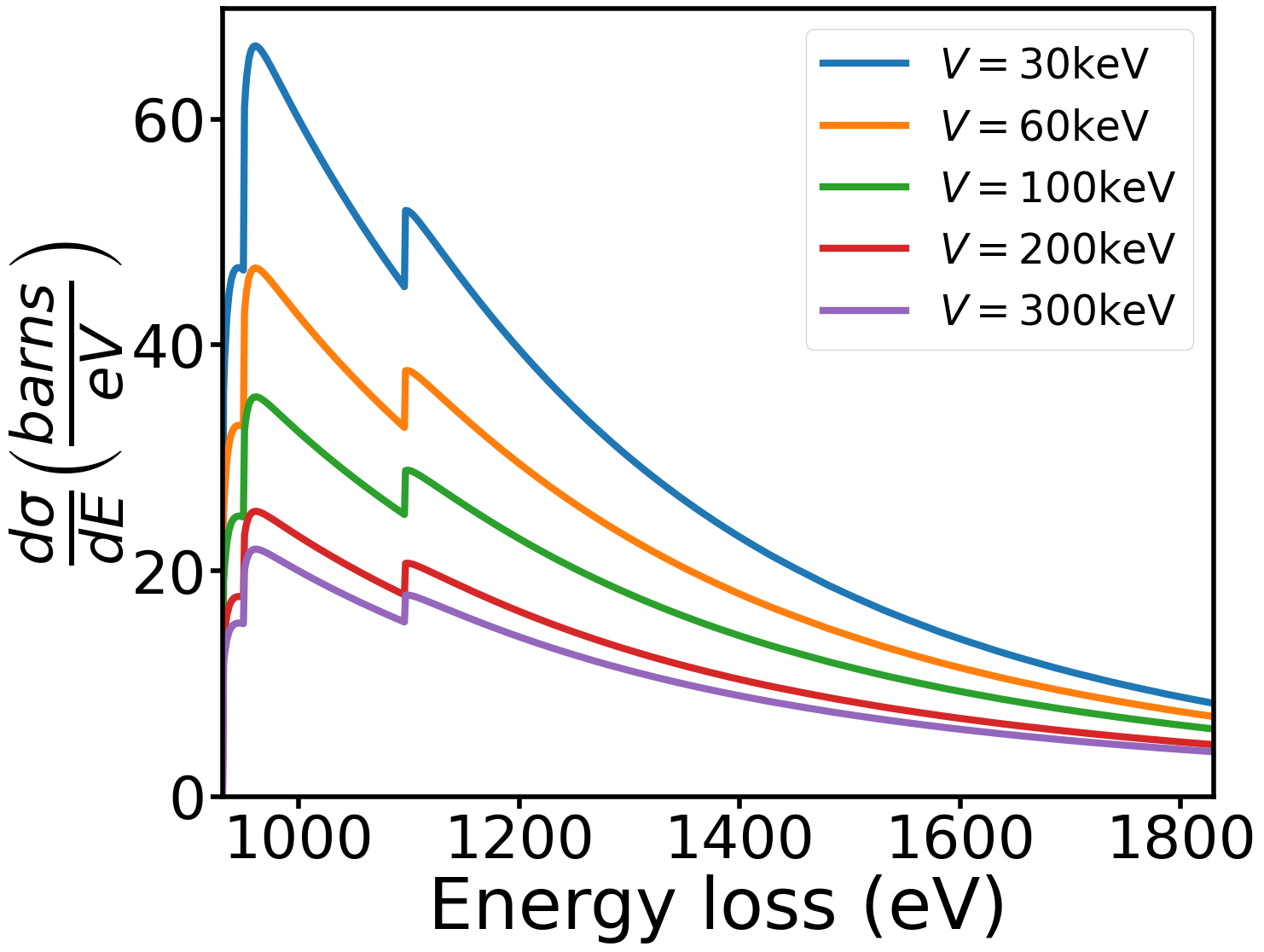}
\end{subfigure}\hspace{5pt}%
\begin{subfigure}[t]{0.9\linewidth}
\caption{}
\includegraphics[width=\linewidth]{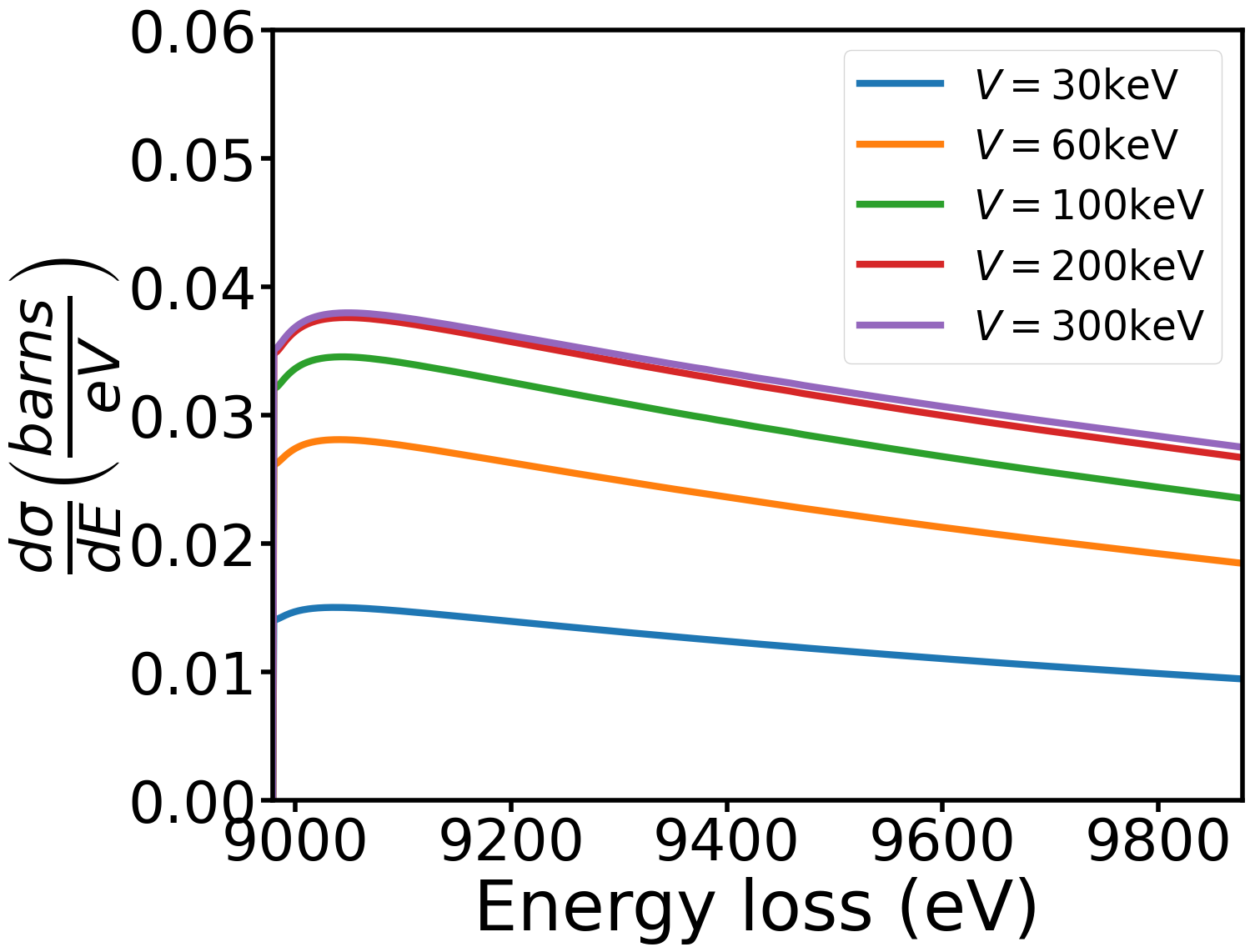}
\end{subfigure}%
\caption{Dependence of the Cu-$L$ (a) and Cu-$K$ (b) differential ionization cross sections, Eq. (\ref{QEDDCSFinal}), on the energy loss of the incoming electron for different acceleration energies of the electron beam. The cross section is calculated accounting for the relaxation of the core hole. The collection angle is $100$ mrad. $V$ denotes the beam acceleration voltage.}
\label{fig:CuKLBeamVoltage}
\end{figure}

\begin{figure}[!htbp] 
\begin{subfigure}[t]{0.8\linewidth}
\caption{}
\includegraphics[width=\linewidth]{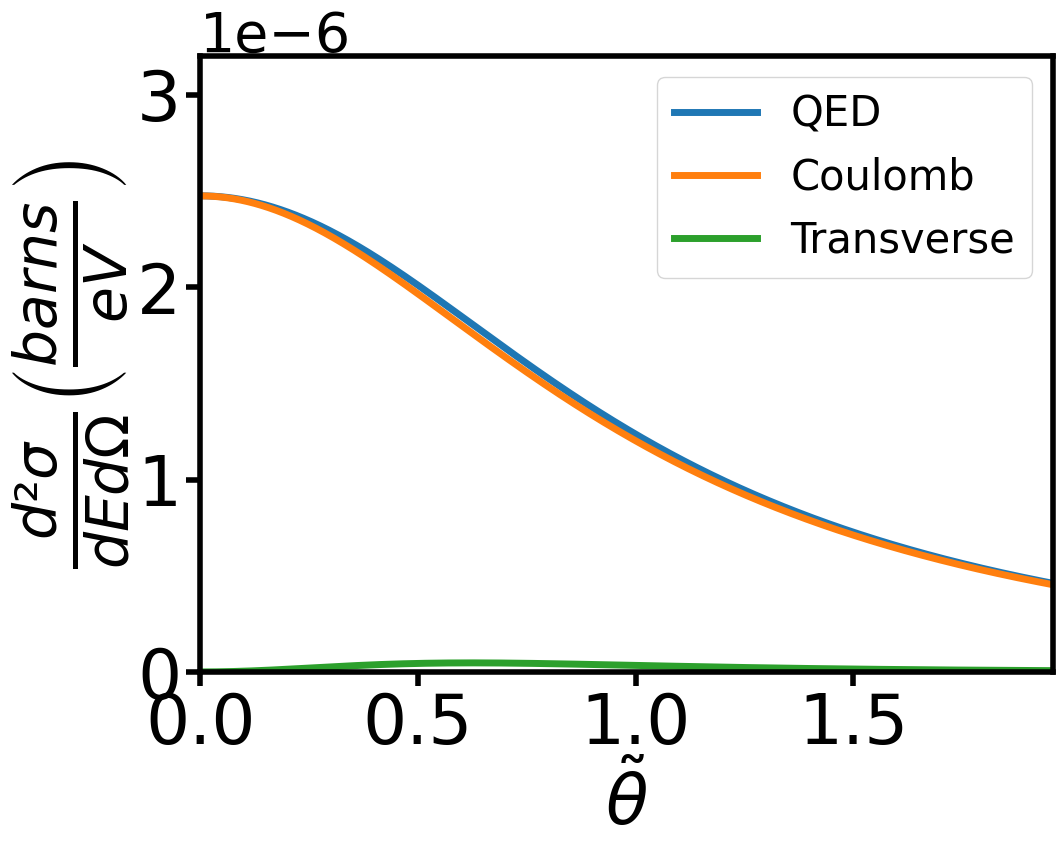}
\end{subfigure}\hspace{5pt}%
\begin{subfigure}[t]{0.85\linewidth}
\caption{}
\includegraphics[width=\linewidth]{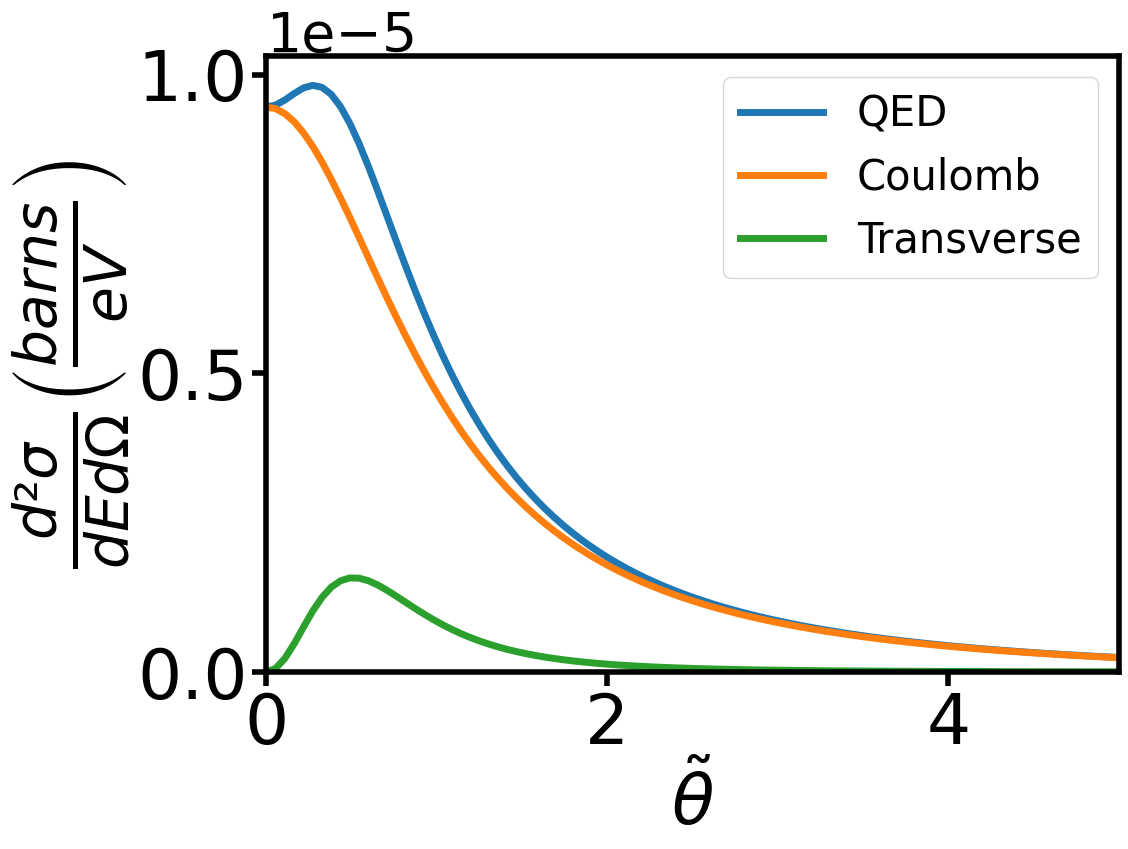}
\end{subfigure}\hspace{5pt}%
\begin{subfigure}[t]{0.85\linewidth}
\caption{}
\includegraphics[width=\linewidth]{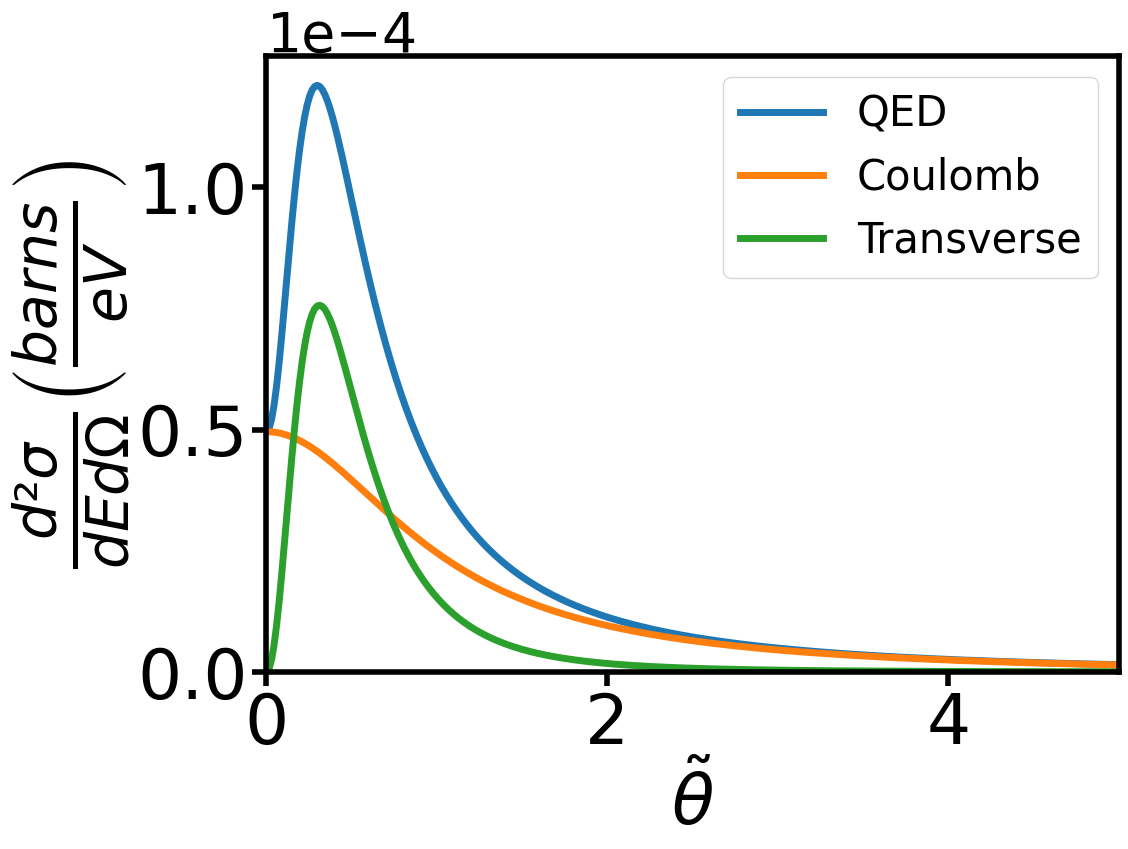}
\end{subfigure}\hspace{5pt}%
\caption{The Cu-$K$ double-differential ionization cross section, Eq.(\ref{QEDDCSFinal}), in terms of the effective collection angle defined in Eq. ~(\ref{tildethetadef}). The beam voltage is $100$ (a), $300$ (b), and $1000$ kV (c). The energy loss is set to the ionization threshold of the Cu-$K$ shell $E_\mathrm{Cu-K} = 8979$ eV.}
\label{fig:CuKSnKDDCS}
\end{figure}

The differential cross section of Cu $K$ and $L$ shells is  also calculated for different acceleration voltages of the incoming electrons, (see Fig. \ref{fig:CuKLBeamVoltage}). The relativistic kinematics of the beam electrons affects the dependence of the cross section on the beam energy.
In case of EELS experiments at high-energy loss or large collection angle, the momentum transfer $|\bfq|$ is beyond the range of validity of the dipole approximation, and the effects of transverse photons become manifest. This makes our fully relativistic result indispensable for accurate analysis. Figure (\ref{fig:CuKSnKDDCS}) shows the contribution of the multipole Coulomb, transverse electric and magnetic transition matrix elements to the double-differential cross section as a function of the effective collection angle $\tilde \theta$, Eq.~(\ref{tildethetadef}). The contribution of the QED part of the matrix element is of the order of $\sim 10\%$. The shape of the double-differential cross section is qualitatively similar to the results found in the literature \cite{Schattschneider2005}, but the actual values differ in the case of large energy loss experiments, as follows directly from the expansion (\ref{dipoleexpansionT}). The numerical differences are small for the majority of low-energy loss EELS experiments.

\begin{figure}[!htbp] 
\includegraphics[width=0.9\linewidth]{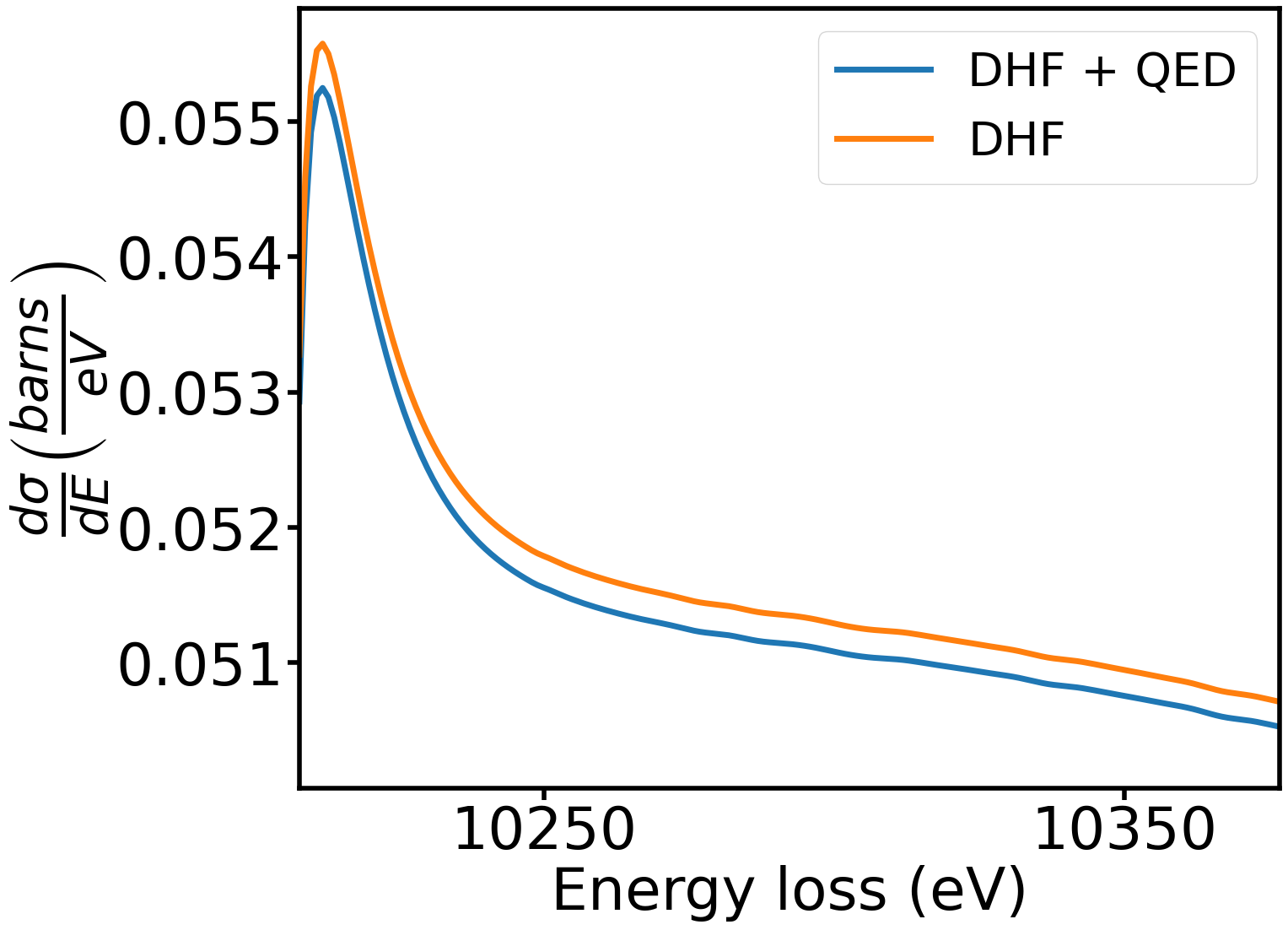}
\caption{The differential ionization cross section, Eq. (\ref{QEDDCSFinal}), for the $\mathrm{W}-L_3$ edge is presented as a function of energy loss in the relaxed core-hole approximation. The atomic structure of tungsten is calculated within DHF with Coulomb electron-electron interaction (orange curve) and with the addition of the transverse QED electron-electron interactions (blue curve) such as the Breit interaction in Eq.~(\ref{eq:breit}). The beam voltage is $300$ kV and the collection angle $100$ mrad.}
\label{fig:WLQED}
\end{figure}

The transverse photon corrections are primarily important in the beam-target interaction, but they also contribute to the dynamics of the deep shells of heavy atoms. These corrections are typically described by effective potentials that are derived from QED. Figure~(\ref{fig:WLQED}) shows the QED differential cross section of $\mathrm{W}-L_3$ subshell in terms of the energy loss. The atomic structure of tungsten is calculated using the Breit and Uehling potential, as well as the electron self-energy terms \cite{ambit, PhysRevA.93.052509, Ginges_2016}. QED corrections on the atomic bound states are small, but they can be important in the analysis of the extended fine structure of high-energy edges in EELS or XAS \cite{PhysRevB.17.560, 10.1093/micmic/ozad067.172}.

\subsection{\label{sec:excDCS} Excitation differential cross section}
The EELS and XAS near-edge fine structure of transition metal and rare-earth elements exhibit rich excitonic behavior due to the strong interaction between the core hole and the valence electrons \cite{GROOT200531, PhysRevB.32.5107}. In this section, we calculate the excitation and ionization differential cross sections of Sc and Dy and compare the results to experimental EELS data. The excitations to discrete states are modeled using atomic multiplet and crystal-field effects as described in \cite{PhysRevB.41.928, PhysRevB.32.5107} and implemented in the \texttt{QUANTY} and \texttt{CRISPY} software packages \cite{Haverkort_2016, retegan_crispy}.
The excitation spectrum of a transition metal and rare-earth elements can be described by local atomic physics because the core hole in the $2p$ or $3d$ orbitals interacts strongly with the localized $3d$ or $4f$ electrons, respectively. The initial and final electron configurations of such systems include open shells, which are appropriately described by $^{2S+1}L_J$ coupled states. The interaction with ligand atoms is modeled using crystal-field theory, where the states are split according to their symmetry representations, which determine the main features of the spectrum. The symmetry group of the $\mathrm{ScO}_3$ lattice is $\mathrm{O}_h$. The ionization part of the differential cross section is calculated within the relaxed DHF framework as described in Sec.~\ref{sec:finalstate}.

\begin{figure}[!htbp] 
\begin{subfigure}[t]{0.8\linewidth}
\caption{}
\includegraphics[width=\linewidth]{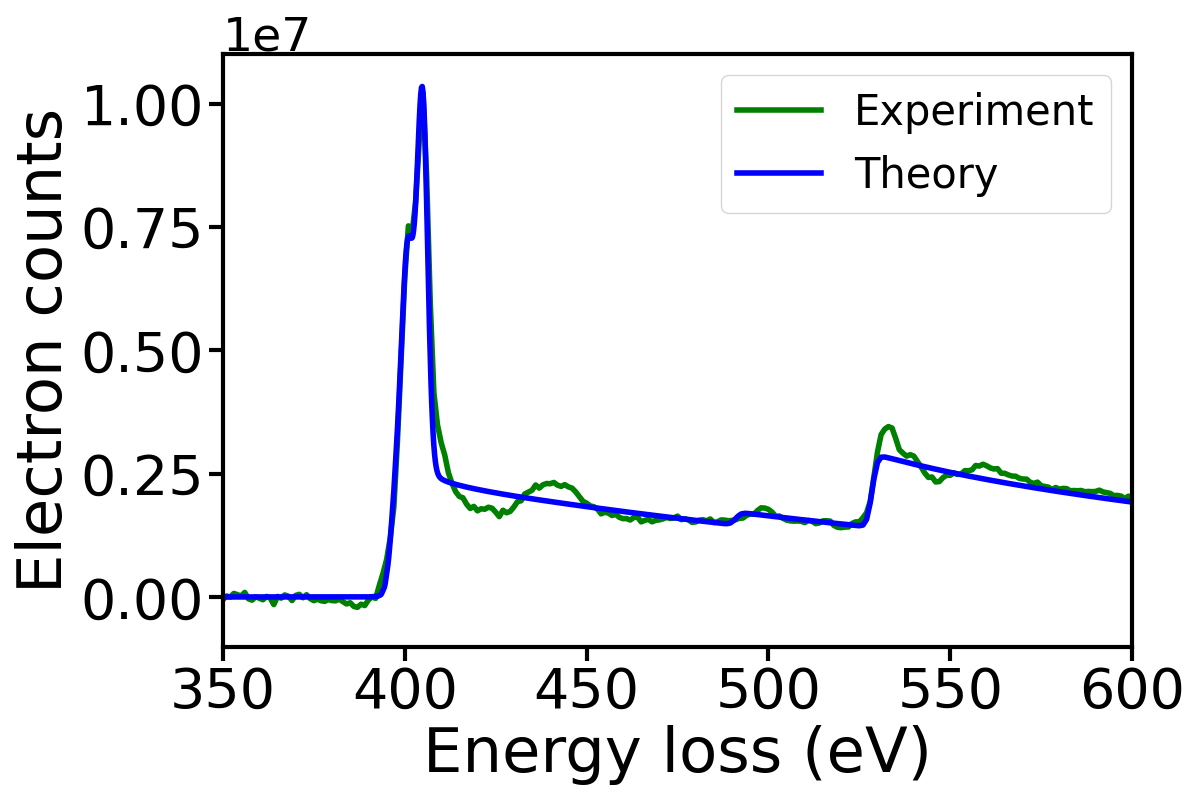}
\end{subfigure}\hfill%
\begin{subfigure}[t]{0.8\linewidth}
\caption{}
\includegraphics[width=\linewidth]{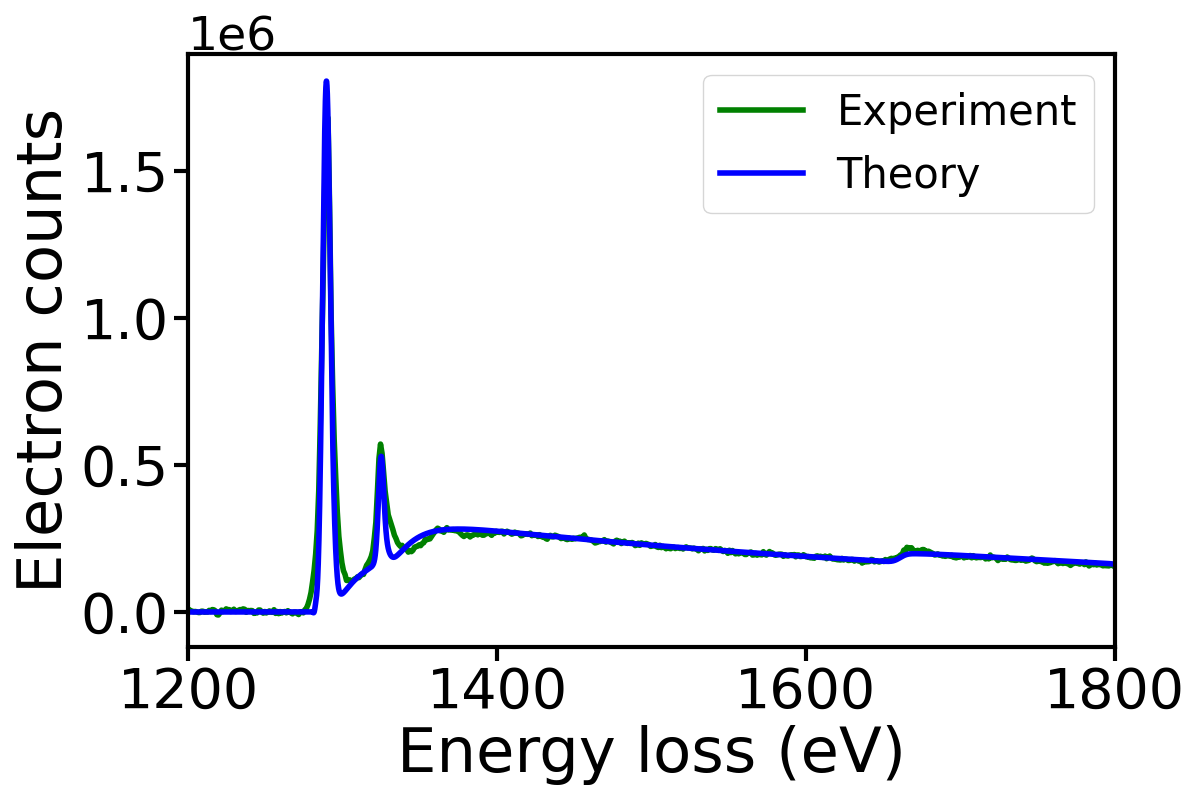}
\end{subfigure}%
\caption{The inelastic differential cross sections of $\mathrm{Sc}-2p$, $\mathrm{O}-1s$ (a) and $\mathrm{Dy}-3d$ (b)  is plotted against the experimental EELS data. The theoretical calculation of Sc-$2p$ and Dy-$3d$ includes the transition to the narrow $3d/4f$ valence band ($2p^63d^0 \to 2p^53d^1$ and $3d^{10}4f^0 \to 3d^94f^1$ respectively) as well as the continuum. The crystal-field parameter for Sc is $10Dq  = 1.7$ eV. The continuum model for all elements corresponds to the continuous spectral function calculated within the relaxed DHF approximation. The beam voltage is 300 kV and the collection angle  40 mrad.}
\label{fig:DyScO3}
\end{figure}

The inelastic cross section, which includes transitions both to the open valence $3d/4f$ shells and to the continuum, is plotted versus an EELS spectrum of $\mathrm{DyScO}_3$ (Fig.~\ref{fig:DyScO3}). We remove the continuous background and fit the broadened theoretical differential cross sections to the Sc, O, and Dy experimental data. For a detailed presentation of the EELS data analysis methodology, see \cite{VERBEECK2004207, Cueva_2012, Van_den_Broek_2023}.  The experiment was carried out with an Iliad Energy Filter at a beam voltage of $300$ kV and relatively low-energy resolution with dispersion of $1 \, \mathrm{eV}$ per channel. The crystal-field potential depends on the crystal-field splitting. By fitting the calculated cross section to the experimental data, we determine both the overall scale factor, dependent on the material composition, and the crystal-field parameter. The total inelastic differential cross section shows excellent agreement with the experimental data. The $3d$ crystal-field splitting of Sc is found to be $1.7$ eV, in close agreement with the value extracted from higher-resolution XAS data \cite{PhysRevB.41.928}. These results indicate that fundamental knowledge of the electronic structure of materials, such as the crystal-field splitting, can be reliably extracted from lower-energy resolution experiments when rigorous theoretical models are employed. Modeling the excitation spectrum of transition metal and lanthanide elements significantly enhances the EELS analysis focused on elemental composition \cite{rezdiscrete}. This approach improves the detection of trace elements and sensitivity to noise, as a substantial portion of the intensity is concentrated in the near-edge region. Additionally, it facilitates improved deconvolution of closely spaced edges, enabling more accurate compositional analysis.

\section{\label{sec:sum}Summary and Discussion}

We carried out a comprehensive analysis of the inclusive inelastic scattering of a relativistic electron beam from a generic heavy target, with results directly applicable to the interpretation of core-loss EELS data. 

The most general form of the relativistic inelastic differential cross section was derived at tree level, expressed in terms of the irreducible tensor operators of the target’s electromagnetic transition current, Eq.~(\ref{QEDDCSFinal}). Such a decomposition allows for the physical classification of the various transition channels contributing to the ionization process. We obtained an analytical expression for the angular dependence of the transition matrix elements, ultimately reducing the problem to simple radial integrals over Dirac orbital wave functions, Eq.~(\ref{Tcoulme}). The resulting formalism is valid across a broad range of experimental conditions, including incident beam energies up to the MeV scale, energy losses up to the order of $O(100~\mathrm{keV})$, arbitrarily large scattering angles and momentum transfer, as all the multipole contributions are accounted for. 

We numerically demonstrated the complete contribution of the transverse photon degrees of freedom  and their impact on the differential cross section of $\mathrm{Cu}-K$ as a function of the effective scattering angle, defined in Eq.~(\ref{tildethetadef}). The relation of the QED ionization differential cross section to the photoionization cross section emerges naturally in this formulation in accordance with the generic structure of QED amplitudes. This connection can be useful for comparing measurements and models across EELS, XAS, and IXS in the relativistic regime beyond the dipole approximation \cite{PhysRevB.58.5989}. The long-wavelength limit of the QED transition matrix element was also derived as a series expansion in terms of the energy loss and the effective scattering angle.

We performed {\it ab initio} atomic structure calculations within the relaxed DHF framework using the \texttt{AMBIT} software \cite{ambit}. The DHF equations of motion were solved rigorously for several elements, incorporating explicit Coulomb and nonlocal exchange interactions among electrons, along with a realistic potential for the electron–nucleus interaction. Quantum electrodynamics (QED) corrections to the electron–electron interaction in the atomic structure, such as the Breit and one-loop terms, were also included. Although these corrections have a non-zero effect, their impact on the orbital wave functions of core shells in heavy atoms was found to be small. Higher-order correlation effects beyond the relaxed DHF method for inner shells have been estimated to provide small contributions in inner-shell ground state properties  \cite{PhysRevA.31.556, ELindroth_1993, indelicato_lindroth_1992, refId0}, and therefore not considered in this work. 

We employed the relaxed DHF framework for the calculation of many-body effects associated with core-shell ionization in atoms. In appendix \ref{app:rpa}, we show that, for core-shell ionization, both the Tamm-Dancoff approximation (TDA) limit of RPAE and relaxed DHF methods yield quantitatively similar corrections, as first demonstrated in \cite{Amusia1971}. The spectral shape of the inner-shell ionization is dominated by the static rearrangement of the atomic orbitals in response to the creation of the core hole. The direct and exchange interactions between the ejected electron and the residual ion are screened by the relaxation of the bound electrons, leading to a reduced differential cross section near the ionization threshold, consistent with previous experimental and theoretical observations in photoionization studies \cite{Amusia1969, AMUSIA1967541, 2006sham.book.379S, AMUSIA1976191}. The range of validity of our approach is determined by the magnitude of the electron-hole overlap integrals and the energy separation of the orbitals which participate in possible virtual excitations. A quantitative estimation of higher-order correlation effects in the ionization cross section of $\mathrm{Ru}-M$ indicates that they contribute less than a few percent relative to the static rearrangement.
The pronounced influence of orbital relaxation on core-loss edge spectral shape has direct consequences for compositional analysis in EELS. In particular, 
$M$ ionization edges, which are difficult to employ in quantitative EELS \cite{HOFER198881}, are strongly modified by core-hole screening, as demonstrated for the $\mathrm{Ru}-M$ and $\mathrm{W}-M$ edges. The methodology developed here provides an accurate yet efficient framework for modeling the common edges for EELS core-loss analysis. We also showed that the explicit relaxation and $Z+1$ approaches for the ionization cross section within DHF are in close agreement. 

The creation of a core hole introduces additional decay channels in the scattering process, such as x-ray emission and Auger decay. These processes can significantly affect the cross section near the ionization threshold, particularly if their lifetimes are comparable to the time scale of the core-hole relaxation. In this work, we approximated the final state as a fully decayed atomic core and analyzed how this approach modifies the shape of the differential ionization cross section. 

Finally, we computed the excitation and ionization spectra for $\mathrm{Sc}-L$ and $\mathrm{Dy}-M$ edges and compared them with experimental EELS data. The excitonic part of the spectrum was calculated using crystal-field multiplet theory, where the open shell structure of the valence states plays a leading role through the LS coupling and the crystal symmetry. Although individual fine-structure peaks are not fully resolved in the experimental data, fitting the crystal-field parameters to the Sc edge yields results in close agreement with previous high-resolution XAS studies \cite{PhysRevB.41.928}. Our calculations provide an improved methodology for elemental quantification of ionic solids via EELS, incorporating the crystal environment of the atom within a controlled theoretical framework.

\subsection{\label{subsec:outlook}Outlook}
Beyond the aspects discussed in this article, several additional directions could further extend this work. Ionization is inherently a dynamic process involving electron density relaxation and secondary processes, such as photon or Auger electron emission, which occur on different timescales. Incorporating core-hole decay effects into the current framework, as suggested in ~\cite{amusia1997inner}, represents a complex yet important step towards a more comprehensive description of simultaneously acquired EELS and EDS data. In particular, the lifetime of excitonic states depends on such processes.

Although the excitation spectra were computed using the crystal-field atomic multiplet method with all final-state electrons fully coupled, the ionization cross section was obtained without accounting for the coupling of the unoccupied discrete and continuum states. This is a reasonable approximation for core-shell ionization in most cases. Incorporating configuration interaction treatment of the discrete-continuum coupling in the calculation of the differential cross section is known to introduce additional broadening and asymmetric peak shapes \cite{PhysRevB.85.165113}. For long-lived excitonic states Fano resonance broadening becomes comparable to the autoionization widths of the excitations. The interaction multiple configurations are required for systems of high covalency where charge transfer influences the spectral shape \cite{GROOT200531}. The example of $\textrm{DyScO}_3$ studied in this work is well described within $LS$ coupling of a single configuration \cite{PhysRevB.41.928, PhysRevB.32.5107}. We emphasize that higher-order correlation corrections, such as Fano mixing and configuration interaction of the ground state, have small impact on experiments aiming at elemental quantification of materials which are typically performed at lower resolution of order $\sim O(1~\mathrm{eV})$. However, these effects may become important for detailed analysis of finer properties such as the oxidation state, spin, or magnetic properties of the material.

Another particularly interesting future direction is the study of lower-energy EELS edges associated with  valence and subvalence subshells. In this case the autoionization processes such as Coster-Kroning and super-Coster-Kroning also introduce additional broadening of the spectral features, especially for excitons close to the ionization threshold as demonstrated in analyses of low-energy transition metal spectra in \cite{SHIN19811281, vanderLaan_1991}. As discussed in Appendix~\ref{app:rpa}, intershell and intrashell correlations have a higher impact in the case of outer shell excitations. As a result, the generalized RPAE in the basis of relaxed final-state orbitals is expected to contribute substantially to the differential cross section \cite{Amusia1971, Amusia_1974}.

Moreover, we considered a high-energy electron beam, where the exchange interaction of the beam with the atomic electrons is neglected since it is suppressed by $\sim 1/{p_i}$. When the incoming electron energy is low enough compared to the ionized electron's kinetic energy, the final state of the entire system should be solved self-consistently. This case can be treated non-relativistically and it is more relevant for low-beam voltage and outer-shell ionization \cite{PhysRevA.38.1279}.

An additional direction for future work is the calculation of the differential-cross section in oriented or anisotropic media. In the present study, we assumed an isotropic distribution of atoms within the target sample. This assumption can be relaxed within the current theoretical framework, enabling analysis of oriented (e.g., magnetic) targets beyond the long-wavelength (dipole) approximation, which has been previously demonstrated in \cite{PhysRevB.59.12807}.

Finally, we note that the QED differential cross section encompasses all contributions to the electron–atom interaction: Coulomb, electric, and magnetic. A detailed comparison of these contributions can offer insights into the kinematic regimes where specific excitation types can be effectively probed. As an example, one can identify the relevant contribution of magnetic excitations, such as magnons, compared to Coulomb excitations as a function of the experimental conditions.

\begin{acknowledgments}
We would like to thank J. Berengut, F. de Groot, D. Sébilleau,  D. Muller, S. Pantelides and M. Haverkort for useful discussions on different aspects of this paper. We are grateful to R. Egoavil for providing the $\text{DyScO}_3$ experimental spectrum. V. Brudanin acknowledges support by an ERC Advanced Grant, under Grant Agreement No. 101055013. 
\end{acknowledgments}

\appendix

\section{\label{ap:feynmandiagram}The Feynman diagram}
The Dyson equation for the $S$-matrix of electron scattering with a composite system is defined in terms of the asymptotically stationary states, 
\cite{Gell-Mann:1953dcn}, 
\cite{VANNIEUWENHUIZEN1967595}.  The $S$-matrix reads as
\begin{equation}
S_{F I} = \langle \Phi^-_F | \Phi^+_I \rangle \, .
\end{equation}
The interaction Hamiltonian is readily derived from the Lagrangian ~(\ref{QEDLagrangian}):
\begin{align}
H_{int}(t) = & -\int d^3 x \, \calj^\m(t, {\bf x}) A_\m(t, {\bf x})
\nn \\  &+ i e \, \int d^3x \,  \bar{\Psi}(t, {\bf x}) \slashed{A}(t, {\bf x}) \Psi(t, {\bf x})
\label{QEDHint}
\end{align}
The initial and final scattering states are the direct product of the incoming electron with the many-particle state of the target $| \Phi \rangle$, $|I\rangle = | p_i m_{s_i} \rangle \otimes |\Phi_\alpha \rangle $  and $ |F \rangle = | p_f m_{s_f} \rangle \otimes | \Phi_\beta \rangle$. respectively. The incoming electron plane wave $\psi_{p, m_s} (x) = e^{ip x}  u(p,m_s)$ is the solution of the free Dirac equation, and the photon propagator in coordinate space reads as
\begin{align}
\Delta_{\mu\nu}(x_1,x_2)= \int {d^4 q \over (2 \pi)^4} {\eta_{\mu\nu} \over q^2 - i \varepsilon} e^{i q(x_1-x_2)} \,.
\end{align}
The photon-fermion vertex is $(2 \pi)^4 e \gamma^{\mu} \delta\left(\sum p\right)$, while the coupling with the external current results in the vertex $(2 \pi)^4 \langle \Phi_\beta | \calj_{\m}(x) | \Phi_\alpha \rangle\ $. The Feynman diagram in Fig.~\ref{feynmandiag}, is readily calculated:
\begin{align}
S_{p_f,m_{s_f},\,\beta \,;\, p_i,m_{s_i},\,\alpha} =&-{e \over (2 \pi)^3} {\bar{u}(p_f,m_{s_f}) \gamma_\m u(p_i, m_{s_i}) \over q^2 -i \varepsilon} 
\nonumber \\ & \times 
\int d^4x e^{-i q x} \langle\Phi_\beta | \calj_{\m}(x) | \Phi_\alpha \rangle\ \,,
\end{align}
From the definition (\ref{Sdef}) the transition matrix is 
\begin{align}
|T_{p_f,\,\beta \,;\, p_i,\,\alpha}|^2  &= {1\over 2} \sum_{m_{s_i},m_{s_f}}   | T_{p_f,m_{s_f},\,\beta \,;\, p_i,m_{s_i},\,\alpha} |^2 
\nonumber \\
&=-{ e^2 \over 2 q^4} {\mathrm Tr}
\left( \gamma^\m {-i {\slashed p_i} + m \over 2 E_i} \gamma^\n {-i {\slashed p_f} + m \over 2 E_f} \right) 
\nonumber \\ & \times 
\langle \Phi_\alpha | \calj_{\n}^{\dagger}(\bfq) | \Phi_\beta \rangle \,
\langle \Phi_\beta | \calj_{\m}(\bfq) | \Phi_\alpha \rangle 
\end{align}
where the matrix $\beta = i \gamma^0$ should not be confused with the target's final-state quantum numbers. The trace is taken over the spinor indices.
The $T$ matrix in terms of the electromagnetic 4-current then becomes
\begin{align}
|T_{p_f,\,\beta \,;\, p_i,\,\alpha}|^2=&-{ e^2 \over 2 q^4} {1 \over E_i E_f}\Big( (p_i \cdot p_f +m^2) |\calj_{\beta\alpha}|^2 
\nonumber \\ &
- p_i \cdot \calj_{\beta\alpha} \, p_f \cdot \calj_{\beta\alpha}^{\dagger}
-p_i \cdot \calj_{\beta\alpha}^{\dagger} \,\, p_f \cdot \calj_{\beta\alpha}   \Big) \,.
\end{align}

\section{\label{app:multipoleexpansion} The multipole expansion derivation for Dirac orbitals}
Equation~(\ref{Tcoulme}) is a general expression of the relevant transition matrix elements in terms of the irreducible tensor operators of the rotation group. Here, we calculate such matrix elements for Dirac wave functions. Similar derivations were performed for photoionization and photon emission of atoms with slightly different methods, see \cite{IPGrant_1974, doi:10.1080/00018737000101191}.
Inserting the expression of the 4-current, Eqs.~(\ref{jdirac0}), (\ref{jdiraci}) in the $T$ matrix elements (\ref{tmatrixJ}) leads to
\begin{widetext}
\begin{align}
\langle \calm || T_{J}^{(mag)}  || \caln \rangle  =& 
 i \,e \,  \left[\langle \k_\calm || \left[ \mathbf{Y}_{J\,J} \cdot \bfs \right] || -\kappa_\caln \rangle
\int dr \, r^2\, j_J\left(\nbfq r\right) F_\calm^\dagger(r) G_\caln (r)
\right. \nonumber \\ 
&\left.
- \langle -\k_\calm || \left[ \mathbf{Y}_{J\,J} \cdot \bfs \right] || \k_\caln \rangle
\int dr \, r^2\, j_J\left(\nbfq r\right) G_\calm^\dagger(r) F_\caln (r)
\right]
\\
\langle \calm || T_{J}^{(el)}  || \caln \rangle = & 
i\, e \, \sqrt{J+1 \over 2 J +1}
\left[
\langle \k_\calm || \left[ \mathbf{Y}_{J\,J-1} \cdot \bfs \right] || -\k_\caln \rangle
\int dr \, r^2\, j_{J-1}\left(\nbfq r\right) F_\calm^\dagger(r) G_\caln (r)
\right . \nonumber \\ &
- \left .
\langle -\k_\calm || \left[ \mathbf{Y}_{J\,J-1} \cdot \bfs \right] || \k_\caln \rangle
\int dr \, r^2\, j_{J-1}\left(\nbfq r\right) G_\calm^\dagger(r) F_\caln (r)
\right]
\nonumber \\ &
-i\, e \, \sqrt{J \over 2 J +1}
\left[
\langle \k_\calm || \left[ \mathbf{Y}_{J\,J+1} \cdot \bfs \right] || -\k_\caln \rangle
\int dr \, r^2\, j_{J+1}\left(\nbfq r\right) F_\calm^\dagger(r) G_\caln (r)
\right . \nonumber \\ &
+ \left .
\langle -\k_\calm || \left[ \mathbf{Y}_{J\,J+1} \cdot \bfs \right] || \k_\caln \rangle
\int dr \, r^2\, j_{J+1}\left(\nbfq r\right) G_\calm^\dagger(r) F_\caln (r)
\right] \,.
\end{align}
In these expressions, we notice the irreducible matrix element of the tensor operator. This matrix element is the inner product of a vector spherical harmonic and the spin operator. This is decomposed into the matrix elements of the operators $Y_{L=J,J\pm1}$ and $\sigma$ that act on the coordinate and spin space, respectively \cite{VMK}:
\begin{equation}
\langle j',\ell',\frac{1}{2}|| \mathbf{Y_{J\,L}} \cdot \bfs ||
j, \ell,\frac{1}{2} \rangle = [j', J, j] \begin{Bmatrix}
\ell' & \ell & L\\
{1\over 2} & {1\over 2} & 1 \\
j' & j & J
\end{Bmatrix} \langle j', \ell' || Y_L || j, \ell \rangle \langle \frac{1}{2} || \sigma || \frac{1}{2} \rangle 
\end{equation}
Applying Eq.~(\ref{Ymatrixelement}) and $ \langle {1 \over 2} || \sigma || {1\over 2} \rangle = \sqrt{6}$ we find
\begin{align}
\langle \calm || T_{J}^{(mag)}  || \caln \rangle  &
= i \, e\, \left[ j_m, \, J, \, j_n \right]^{1\over 2} \langle {1 \over 2} \vert \vert \sigma || {1\over 2} \rangle
\nonumber \\
&\left[
\langle \kappa_\calm || Y_{J} || -\kappa_\caln \rangle
\begin{Bmatrix}
\ell_\calm^- & \ell_\caln^+ & J\\
{1\over 2} & {1\over 2} & 1 \\
j_\calm & j_\caln & J
\end{Bmatrix}
\int dr \, r^2\, j_J\left(\nbfq r\right) F_\calm^\dagger(r) G_\caln (r)
\right .
\nonumber \\
&\left .
-\langle -\kappa_\calm || Y_{J} || \kappa_\caln \rangle
\begin{Bmatrix}
\ell_\calm^+ & \ell_\caln^- & J\\
{1\over 2} & {1\over 2} & 1 \\
j_\calm & j_\caln & J
\end{Bmatrix}
\int dr \, r^2\, j_J\left(\nbfq r\right) G_\calm^\dagger(r) F_\caln (r) 
\right]
\nonumber \\
& =
i \, e\, (-1)^{\ell^-_\calm} \sqrt{\left[ j_\calm, \, J, \, j_\caln \right] \over 4 \pi}
\Pi(\ell^-_\calm, J, \ell^+_\caln)
\nonumber \\ &\times 
\left[
\sqrt{6 \, [\ell^-_\calm, \, J, \, \ell^+_\caln ] } 
\begin{pmatrix}
\ell^-_\calm & J & \ell^+_\caln \\
0 & 0 & 0
\end{pmatrix}
\begin{Bmatrix}
\ell^-_\calm & \ell^+_\caln & J\\
{1\over 2} & {1\over 2} & 1 \\
j_\calm & j_\caln & J
\end{Bmatrix} 
\right.
\int dr \, r^2\, j_J\left(\nbfq r\right) F_\calm(r) G_\caln (r)
\nonumber \\ &
\left.
+\sqrt{6 \, [\ell^+_\calm, \, J, \, \ell^-_\caln ]} 
\begin{pmatrix}
\ell^+_\calm & J & \ell^-_\caln \\
0 & 0 & 0
\end{pmatrix}
\begin{Bmatrix}
\ell^+_\calm & \ell^-_\caln & J\\
{1\over 2} & {1\over 2} & 1 \\
j_\calm & j_\caln & J
\end{Bmatrix}
\int dr \, r^2\, j_J\left(\nbfq r\right) G_\calm(r) F_\caln (r) 
\right]
\nonumber \\
=&-i \, e\, (-1)^{j_\calm+1/2} \sqrt{\left[ j_\calm, \, J, \, j_n \right] \over 4 \pi}
\begin{pmatrix}
j_\calm & j_\caln & J \\
-{1\over 2} & {1 \over 2} & 0
\end{pmatrix}
\Pi(\ell^-_\calm, J, \ell^+_\caln)
{(\k_\caln+\k_\calm) \over \sqrt{J(J+1)}}
\nonumber \\ & \times
\int dr \, r^2\, j_J\left(\nbfq r\right) 
\left( F_\calm(r) G_\caln(r)+G_\calm(r) F_\caln (r) \right)
\end{align}
where $[x,y,\ldots] = (2x+1)(2y+1)\ldots$ and $\Pi(\ell^-_\calm, J, \ell^+_\caln) =1 $ for $\ell^-_\calm+J+\ell^+_\caln=2 k$, ($k \in \mathbb{Z}$) and $0$ otherwise. Notice $\ell^{\pm} = J \mp {\mathrm{sgn}(\kappa) \over 2}$ and $J=\mathrm{sgn}(\kappa) \, \kappa -{1\over2}$. The following identity was applied for the reduction of 9-$j$ Wigner symbols to 3-$j$ Wigner symbols \cite{10.1143/PTP.11.143}: 

\begin{align}
&
\sqrt{6 \, [J, \, \ell_\calm, \, \ell_\caln ]} 
\begin{pmatrix}
J & \ell_\calm & \ell_\caln \\
0 & 0 & 0
\end{pmatrix}
\begin{Bmatrix}
\ell_\calm & \ell_\caln & J\\
{1\over 2} & {1\over 2} & 1 \\
j_\calm & j_\caln & J
\end{Bmatrix}
=
\begin{pmatrix}
j_\calm & j_\caln & J \\
{1\over 2} & {1 \over 2} & -1
\end{pmatrix}.
\end{align}
The electric $T$ matrix reads as
\begin{align}
\langle \calm || T_{J}^{(el)}  || \caln \rangle 
& =
i \, e\, \left[ j_\calm, \, J, \, j_\caln \right]^{1\over 2} \langle {1 \over 2} || \sigma || {1\over 2} \rangle
\nonumber \\ &
\left[
\sqrt{J+1 \over 2 J +1} \, \langle \k_\calm || Y_{J-1} || -\k_\caln  \rangle
\begin{Bmatrix}
\ell^-_\calm  & \ell^+_\caln  & J-1\\
{1\over 2} & {1\over 2} & 1 \\
j_\calm & j_\caln & J
\end{Bmatrix}
\right . 
\int dr \, r^2\, j_{J-1}\left(\nbfq r\right) F_\calm^\dagger(r) G_\caln(r)
\nonumber \\
&
-\sqrt{J+1 \over 2 J +1} \,\langle -\k_\calm || Y_{J-1} || \k_\caln  \rangle
\begin{Bmatrix}
\ell^+_\calm & \ell^-_\caln  & J-1\\
{1\over 2} & {1\over 2} & 1 \\
j_\calm & j_\caln & J
\end{Bmatrix}
\int dr \, r^2\, j_{J-1}\left(\nbfq r\right) G_\calm^\dagger(r) F_\caln (r) 
\nonumber \\
& 
-\sqrt{J \over 2 J +1} \, \, \langle \k_\calm || Y_{J+1} || -\k_\caln \rangle
\begin{Bmatrix}
\ell^-_\calm & \ell^+_\caln & J+1\\
{1\over 2} & {1\over 2} & 1 \\
j_\calm & j_\caln & J
\end{Bmatrix}
\int dr \, r^2\, j_{J+1}\left(\nbfq r\right) F_\calm^\dagger(r) G_\caln (r)
\nonumber \\ &
\left.
+\sqrt{J \over 2 J +1} \, \langle -\k_\calm || Y_{J+1} || \k_\caln \rangle
\begin{Bmatrix}
\ell^+_\calm & \ell^-_\caln & J+1\\
{1\over 2} & {1\over 2} & 1 \\
j_\calm & j_\caln & J
\end{Bmatrix}
\int dr \, r^2\, j_{J+1}\left(\nbfq r\right) G_\calm^\dagger(r) F_\caln (r) 
\right] \,.
\end{align}
The 9-$j$ symbols are reduced using the identity \cite{10.1143/PTP.11.143}  
\begin{align}
&
\sqrt{6 \, [J \pm 1, \, \ell_\calm, \, \ell_\caln ]}
\begin{pmatrix}
J\pm 1 & \ell_\calm & \ell_\caln \\
0 & 0 & 0
\end{pmatrix}
\begin{Bmatrix}
\ell_\calm & \ell_\caln & J \pm 1\\
{1\over 2} & {1\over 2} & 1 \\
j_\calm & j_\caln & J
\end{Bmatrix}
=
\frac{(\ell_\calm -j_\calm)(2 j_\calm+1)+(\ell_\caln-j_\caln)(2j_\caln+1) \pm J \pm \frac{1}{2} + \frac{1}{2}}{(2 J +1) (J\pm \frac{1}{2} -\frac{1}{2} )}
\begin{pmatrix}
j_\calm & j_\caln & J \\
{1\over 2} & {1 \over 2} & -1
\end{pmatrix}\,.
\end{align}
The electric matrix element then reduces to

\begin{align}
\langle \calm || T_{J}^{(el)}  || \caln \rangle =&
-i \, e\, (-1)^{j_\calm+1/2}
 \sqrt{\left[ j_\calm, \, J, \, j_\caln \right] \over 4 \pi }
 \begin{pmatrix}
j_\calm & j_\caln & J \\
-{1\over 2} & {1 \over 2} & 0
\end{pmatrix}
\Pi(\ell^-_\calm, J, \ell^-_\caln)
\nonumber \\ & 
\times \left[
{\kappa_\caln - \kappa_\calm \over \sqrt{J (J+1)}}
\int dr \, r^2\, { (J+1) j_{J-1}\left(\nbfq r\right)- J j_{J+1}\left(\nbfq r\right)  \over 2 J +1} 
\left( F_\calm (r)G_\caln(r)+G_\calm(r) F_\caln (r)\right)
\right.
\nonumber \\
& 
\left.
+\sqrt{J (J+1)}
\int dr \, r^2\, { j_{J-1}\left(\nbfq r\right)+ j_{J+1}\left(\nbfq r\right)  \over 2 J +1} 
\left( F_\calm (r)G_\caln(r)-G_\calm(r) F_\caln (r) \right)
\right]\,.
\end{align}
\end{widetext}

\section{\label{app:matrixelement}  General multielectron matrix elements}

Here, we calculate the matrix element of a generic one-body operator between two multielectron wave functions, $| \Phi_\a \ra$ and $| \Phi_\b \ra$, respectively. Since the atomic orbitals rearrange themselves after the ionization of the system, the initial- and final-state orbitals are not orthogonal. This makes the relevant transition matrix elements slightly more complicated \cite{Shirley}. An explicit derivation for fully antisymmetric wave functions, written as Slater determinants, is presented below.

Consider a one-body operator
\begin{equation}
{\it O} = \sum_{\widetilde{\calm},\, \caln} \langle \widetilde{\calm} | {\it O} | \caln \rangle a^{\dagger}_{\widetilde \calm} a_{\caln}
\label{onebodyop}
\end{equation}
in two orthonormal bases ${ | \caln \rangle }$ and ${|\widetilde{\calm} \rangle}$.
We calculate the matrix element $ \langle \Phi_\beta | {\it O} | \Phi_\alpha \rangle$, where the initial and final states are Slater determinants in different bases ${ | \caln \rangle }$ and ${|\widetilde{\calm} \rangle}$, respectively. The creation operator of a state $\chi$ in an $N$ particle state acts like
\begin{equation}
a^{\dagger}_{\chi} | \caln_1 \ldots \caln_N\rangle = | \chi \caln_1 \ldots \caln_N\rangle
\end{equation}
By taking the Hermitian conjugate of the equation above and multiplying with an arbitrary ($N-1$) bra state on the left one can show
\begin{equation}
a_{\chi} | \caln_1 \ldots \caln_N\rangle = \sum_k (-1)^{k-1} \langle \chi | \caln_k \rangle \, | \caln_1 \ldots \caln_{k-1} \caln_{k+1} \ldots \caln_N\rangle
\end{equation}
This leads to the matrix element between two $N$-particle states
\begin{align}
\langle \Phi_\b | {\it O} | \Phi_\a \rangle &=
\langle \widetilde{\calm}_1 \ldots \widetilde{\calm}_N | {\it O}| \caln_1 \ldots \caln_N\rangle 
\nonumber \\ &
= \sum_{i, j} \la \widetilde{\calm}_i | {\it O} |  \caln_j \ra (-1)^{i+j} S_{ij} \,,
\label{fullTM}
\end{align}
where
\begin{equation}
S_{ij}=\sum_{P} (-1)^P \prod\limits_{\substack{k=1 \\ k\neq i\,,P(k)\neq j}}^N \la \widetilde{\calm}_k | \caln_{P(k)}\ra
\end{equation}

\section{\label{app:corehole} The core-hole perturbation}

We calculate the perturbative correction of the target's final state due to the core hole ($\call$) and show how it modifies the matrix element of a one-body excitation operator. We show that this matches the variational approach we employed for the calculation of the transition matrix elements. We follow the general principles of the approach outlined in \cite{Amusia1971}.

\begin{figure*}[!htbp]
\begin{minipage}[t]{0.18\linewidth}
\includegraphics[height=0.95\linewidth]{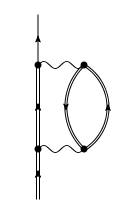}
\caption{The polarization diagram of the vacancy shell.}
\label{fig:MBPol}
\end{minipage}\hfill%
\begin{minipage}[t]{0.78\linewidth}
\includegraphics[width=.99\linewidth]{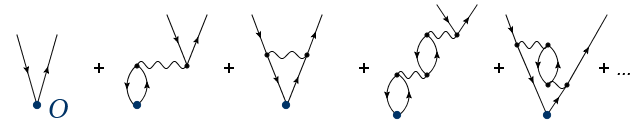}
\caption{Diagrammatic expansion of the matrix element of the operator $\it{O}$ between an electron and a hole state. The electron-hole pair interacts with  the atomic medium  via the direct Coulomb and exchange interaction (wavy lines). The opposite arrows refer to the propagation of an electron and a hole.}
\label{fig:MBRPA}
\end{minipage}%
\end{figure*}

The DHF equations of motion in the presence of a core-hole potential are given in Eq.~(\ref{dhfcorehole}), where the contribution of the electron in the hole shell has been removed from Eqs.~(\ref{cordeom}).
When $\caln \le F$, the back-reaction of the ionized electron to the bound states is not considered. Using the completeness relation of the unperturbed atomic orbitals, Eq.~(\ref{dhfcorehole}) becomes

\begin{flalign}
    \widetilde{H}_{DHF}\, \widetilde{\psi}_\caln(\bfx) 
    -\sum_{\calm > F} \psi_\calm(\bfx) \la {\widetilde \call} \, \calm | V_{ee} | \widetilde{\call} \, \widetilde{\caln} \ra 
    = \tilde{\epsilon}_{\caln} \widetilde{\psi}_\caln(\bfx) 
\end{flalign}
where the orthogonality of the full eigenstates to the unperturbed ones is imposed by restricting the sum over $\calm$ to the unoccupied states. The matrix element of the two-body interaction between orbitals $| \caln \rangle$ and $| \calm \rangle$ includes the direct and exchange terms
\begin{align}
    & \la \tilde{\call}\, \calm | V_{ee} | \tilde{\call} \, \tilde{\caln}\ra 
     =
     \nn \\ &
     \int d\bfx'' \int d\bfx'
    \widetilde{\psi}^{\dagger}_{\call}(\bfx') \psi^{\dagger}_{\calm}(\bfx'') V_{ee}(|\bfx - \bfx'|) \widetilde{\psi}_{[\call} (\bfx') \widetilde{\psi}_{\caln]}(\bfx'') 
 \label{eepotential}
\end{align}
The equation of motion can be rewritten in terms of the unperturbed wave functions
\begin{widetext}    
\begin{align}
    H_{DHF}\, \widetilde{\psi}_\caln(\bfx) + 
    \sum_{\calm > F} \psi_\calm(\bfx) 
    \left(  \sum_{\mathfrak{j} \le F,\, \mathfrak{j} \neq \call}  
    \la \widetilde{\frakj} \, \calm | V_{ee} | \widetilde{\frakj} \, \widetilde{\caln} \ra 
    - \sum_{\frakj \le F} \la \mathfrak{j} \, \calm | V_{ee} | \frakj \, \widetilde{\caln} \ra 
    \right)
    = \tilde{\epsilon}_{\caln} \widetilde{\psi}_\caln(\bfx) \,.
\end{align}
We consider that an ejected electron at state $\caln>F$ does not back-react to the bound-state orbitals. To the lowest order,  the parentheses above are approximated by 
\begin{align}
\left(  \sum_{\frakj \le F,\, \frakj \neq \call}  
    \la \widetilde{\frakj} \, \calm | V_{ee} | \widetilde{\frakj} \, \widetilde{\caln} \ra 
    - \sum_{\mathfrak{j} \le F} \la \mathfrak{j} \, \calm | V_{ee} | \frakj \, \widetilde{\caln} \ra 
    \right) \approx \la \call\, \calm | V_{ee} | \call \, \tilde{\caln}\ra  
\end{align}
The perturbed wave function then reads as
\begin{align}
    | \widetilde{\caln} \ra & =| \caln \rangle + \sum_{\calm>F}\frac{\la \call \, \calm | V_{ee}| \call \, \widetilde{\caln}\ra}{\e_\calm - \e_\caln + i \eta } | \calm \ra
    =| \caln \rangle + \sum_{\calm>F}\frac{ \la \call \, \calm | V_{ee}| \call \, \caln\ra}{\e_{\calm} - \e_\caln + i \eta } | \calm \ra
    +\sum_{\calm,\, \calm'>F}\frac{\la \call \, \calm | V_{ee}| \call \, \calm'\ra \la \call \, \calm' | V_{ee}| \call \, \caln\ra}{(\e_{\calm'} - \e_\caln + i \eta)(\e_\calm - \e_\caln + i \eta) } | \calm \ra + \ldots
\end{align}
Equation~(\ref{fullTM}) can be expanded in the core-hole potential

\begin{align}
    \la \Phi_\b | {\it O} | \Phi_\a \ra = & 
    \sum_{i\, j} (-1)^{i+j}
    \left( \la \caln_i | {\it O} | \caln_j \ra + \sum_{\calm_i>F} \la \calm_i |{\it O} | \caln_j \ra \frac{ \la \call \, \caln_i | V_{ee}| \call \, \calm_i\ra}{\e_{\calm} - \e_\caln + i \eta } +\ldots \right)
    \nn \\ & 
    \times \sum_{P} (-1)^P \prod\limits_{\substack{k=1 \\ k\neq i\,,P(k)\neq j}}^N \left( \la \caln_k | \caln_{P(k)}\ra
    +\sum_{\calm_k>F} \la \calm_k | \caln_{P(k)} \ra \frac{ \la \call \, \caln_k | V_{ee}| \call \, \calm_k\ra}{\e_{\calm} - \e_\caln + i \eta } + \ldots
    \right)
    \nn \\
    \simeq &
    \sum_{i\, j} (-1)^{i+j}
    \left( \la \caln_i | {\it O} | \caln_j \ra + \sum_{\calm_i>F} \la \calm_i |{\it O} | \caln_j \ra \frac{ \la \call \, \caln_i | V_{ee}| \call \, \calm_i\ra}{\e_{\calm} - \e_\caln + i \eta } + \ldots \right)
    \nn \\ &
    \times \left( 1-
    \sum_{p<q \,, p,q \neq i,j} \frac{ \left| \la \call \, \caln_p | V_{ee}| \call \, \caln_q\ra \right|^2}{(\e_{\caln_p} - \e_{\caln_q})^2 } +\ldots
    \right)
    \label{MBMatrixElement}
\end{align}
\end{widetext}
The last parentheses correspond to the vacancy diagonal self-energy correction beyond the mean-field part. The self-energy correction can also be derived from Dyson's equation  by setting the vertex potential equal to the electron-electron potential that leads to the first correction beyond the mean-field approach. The corresponding diagram is shown in Fig.~\ref{fig:MBPol}, where the double lines correspond to the full vacancy propagator in the many-particle medium. The self-energy shifts the ionization potential of the atom. It has been shown to be an important contribution in {\it ab initio} calculation of the ionization and x-ray emission energies of atoms \cite{indelicato_lindroth_1992}. 

The first parentheses correspond to the matrix element of $\it{O}$ expanded in a series of corrections arising from the core-hole potential. The first term is the matrix element of the transition between the unperturbed orbitals $\caln_j$ and $\caln_i$, as shown in the first diagram of Fig.~\ref{fig:MBRPA}, where the operator $\it{O}$ creates the electron-hole pair that moves freely in the medium. The second term corresponds to the creation of an electron-hole pair which interacts with the medium via the electron-electron potential. It is readily noticed that the mixing of the core hole with unoccupied states is suppressed by one over their energy difference. The second and third diagrams in Fig.~\ref{fig:MBRPA} correspond to the direct and the exchange interactions of the electron-hole pair, respectively.  The last two diagrams in Fig.~\ref{fig:MBRPA} correspond to second-order corrections. We notice that the particle-hole diagrams correspond to the time-forward RPAE diagrams as it is further explained in appendix \ref{app:rpa}. However, the electron-hole potential is not weak since it is induced by a localized inner-shell vacancy. The variational framework appropriately addresses the static, nonperturbative response of the atomic medium to the creation of the core hole. This embodies the core-hole effect on the orbitals beyond linear response as it is shown in Eq.~(\ref{MBMatrixElement}).

\subsection{\label{app:rpa} Random phase approximation with exchange}
In the previous section, we summarized how the DHF solution for the state of a system with a core hole is related to the diagrammatic expansion of the self-energy and the polarization propagator. Here, we will start from the RPAE equations and show that they reduce to the Hartree-Fock equations of motion with a core hole \cite{Amusia1971, PhysRevA.74.062503}.

The excitation spectrum of a many-body system, which experiences an external perturbation of the type (\ref{onebodyop}), is described by the polarization propagator, i.e., the four-point Green's function
\begin{align}
    i &\Pi_{\calm \call; \calm' \call'}(t-t')  =
    \nn \\ 
    &\left\langle \Phi_\alpha \right| \mathcal{T} \left( a_{\call}^{(H) \, \dagger}(t) a_{\calm}^{(H)}(t) a_{\calm'}^{(H) \, \dagger}(t') a_{\call'}^{(H)}(t') \right) \left|  \Phi_\alpha \right\rangle \, .
    \label{4ptfunction}
\end{align}
The polarization propagator describes the creation, propagation, and interaction of a particle-hole pair in a many-body medium. The single-particle orbitals here correspond to the DHF orbitals. The RPAE approximation for the correlation function implies that a particle-hole pair interacts with the medium through the Coulomb and exchange potential which is taken to be the vertex function in Dyson's equation for the polarization propagator

\begin{align}
    & \Pi^{(RPA)}_{\calm \call; \calm' \call'}(E) = \Pi^{(0)}_{\calm \call; \calm' \call'}(E)
    \nn \\
    &+\sum_{\calm_1\, \calm_2 \, \call_1 \, \call_2} \Pi^{(0)}_{m \call; \calm_2 l_2}(E) \langle \calm_2 \call_2 | V_{ee} | \calm_1 \call_1 \rangle \Pi^{(RPA)}_{\calm_1 \call_1 ; \calm' \call'}(E)
\end{align}
We define the particle-hole amplitudes 
\begin{align}
    X_{\calm \call} & \equiv \langle \Phi_\alpha | a^\dagger_\call a_\calm | \Phi_\beta \rangle
    \nonumber \\
    Y_{\calm \call} & \equiv \langle \Phi_\beta | a^\dagger_\calm a_\call | \Phi_\alpha \rangle \, ,
\end{align}
Substituting to Dyson's equation the Lehman representation of the Green's function (\ref{4ptfunction}),  we find the well-known equations of motion for the amplitudes

\begin{align}
    (&\epsilon_\calm-\epsilon_\call- E) X_{\calm \call} + \sum_{\calm' \call'} \Big( \langle \calm \call' \left| V_{ee} \right|  \call \calm' \rangle  X_{\calm' \call'} 
    \nn \\
    &+ \langle \calm \calm' \left| V_{ee} \right|  \call \call' \rangle  Y_{\caln \call'} \Big) = 0
    \label{eq:rpa1}
    \\
    (&\epsilon_\calm-\epsilon_\call+ E) Y_{\calm \call} + \sum_{\calm' \call'} \Big( \langle \calm' \call \left| V_{ee} \right|  \call' \calm \rangle  Y_{\calm' \call'} 
    \nn \\
    & + \langle \call \call' \left| V_{ee} \right|  \calm \calm' \rangle  X_{\calm' \call'} \Big) = 0 
    \label{eq:rpa2}
\end{align}
For a detailed derivation, we refer to \cite{fetter2003quantum}. When the particle-hole interaction is neglected, the above equations are solved by the one-particle transition amplitudes. In this case, energy conservation guarantees the solution of the above equations since $E = \epsilon_\calm - \epsilon_\call$. However, the creation of the particle-hole pair induces the Coulomb and exchange potentials, which result in two additional terms.

In case of an excitation or ionization an inne -shell 
with large binding energy close to the unperturbed particle-hole energy, i.e.,  $E \sim \epsilon_\calm - \epsilon_\call$,  Eq. (\ref{eq:rpa1}) decouples from (\ref{eq:rpa2}). This limit corresponds to the Tamm-Dancoff approximation which contains the time-forward diagrams of RPAE with one electron-hole pair at any time \cite{fetter2003quantum}. The energy of the orbital $\calm$ is replaced from the DHF equations of motion in Eq. (\ref{eq:rpa1}),

\begin{align}
    \Big(&-\delta_{\calm \calm'} \epsilon_\call + \langle \calm \left| (H_D + V_{nucl}) \right|  \calm' \rangle 
    \nn\\
    &+ \sum_{ \mathfrak{l'} \neq \call} \langle \calm \mathfrak{l'} \left| V_{ee} \right| \calm' \mathfrak{l'} \rangle  \Big) X_{\calm \call}  = E X_{\calm \call} 
\end{align}
where $H_D$ here denotes the Dirac Hamiltonian.
This matrix equation corresponds to the DHF equations of motion of an ion with a vacancy in the orbital $\call$. This exactly corresponds to the final-state approximation where there is always one electron-hole pair propagating forward in time. The final state of the target is calculated by explicitly including the core hole in the final atomic configuration and the electron-hole potential includes the Coulomb and exchange interaction. 

Inner-shell ionization is accurately described in the above scheme since the absorbed energy from the atomic target is large and close to the energy difference of the orbitals. The TDA approximation is valid in the limit where the backward-propagating terms in Eqs. (\ref{eq:rpa1}, \ref{eq:rpa2}) are small

\begin{equation}
{{\langle \calm \calm' | V_{ee}| \call \call'  \rangle} \over E+(\epsilon_{\calm} -\epsilon_{\call})} \ll 1
\,.
\label{tdalimit}
\end{equation}
For a core hole ionization the energy loss $E$ and the absolute eigenenergy of the vacancy orbital $|\epsilon_{\call}|$ are typically higher than $100\, \mathrm{eV}$. Hence, close to the ionization threshold, the denominator of the above expression is approximately  $\sim 2 |\epsilon_{\call}| \sim O(100~\, \mathrm{eV})$, which suppresses the backward-propagating correlation contributions to the ionization process. Equivalently, the Lehman representation of Eq.(\ref{4ptfunction}) makes manifest that the polarization propagator is dominated by the forward-propagating electron-hole contribution close to the ionization threshold \cite{fetter2003quantum}. 

Intershell and intrashell correlation effects are incorporated in the the full RPAE scheme through the contribution of virtual electron-hole excitations. In \cite{Amusia1971} the RPAE contributions to  the transition matrix elements in photoionization were derived. The authors showed that for a transition from orbital $\call$ to $\calm$ by an operator $\it{O}$, the RPAE matrix element, including the virtual electron-hole pair ($\call'-\calm'$) contributions, is of the form
\begin{align}
\la \calm | {\it O} | \call \ra + \sum_{\calm' > \mathrm{F} ,\, \call' < \mathrm{F}}
\Bigg[ &\frac{\la \calm' | {\it O} | \call' \ra \langle \calm \call' | V_{ee}| \call \calm' \rangle}{E-(\epsilon_{\calm'} -\epsilon_{\call'})}
\nn \\
&- 
\frac{\la \call' | {\it O} | \calm' \ra \langle \calm \calm' | V_{ee}| \call \call' \rangle}{E+(\epsilon_{\calm'} -\epsilon_{\call'})}\Bigg]
\label{correlations}    
\end{align}
For inner-shell ionization, terms such as the first one in the sum are small because the overlap of the inner $\call$ and outer $\call'$ orbitals is small compared to the matrix element of ${\langle \calm \call | V_{ee}| \call \calm' \rangle}$. This matrix element corresponds to a fixed core hole in the static rearrangement approximation.
Backward-propagating terms, such as the last one in Eq.~(\ref{correlations}), are suppressed in the TDA limit, close to the ionization threshold as it was discussed above.  When the hole propagates to a higher-energy shell $\call'$ the excited shell corresponds to similar $E$ as the real electron-hole pair 
when it has energy $\epsilon_{\calm'} \sim \epsilon_{\calm} -  \epsilon_{\call}+ \epsilon_{\call'}$. Then the real and virtual pair contributions have similar denominators. For a core-hole orbital $\call$,  the energy difference of $\call$ and $\call'$ is large and their overlap integral will be small leading to a subleading intershell contribution. This qualitative argument justifies the dominating contribution of static versus dynamic correlation effects in inner-shell ionization. 

In case of Ru-$3d$ ionization, we estimated the ratio of the static rearrangement contribution to the Coulomb transition  with the intershell forward and backward contributions with  respect to $4s$ subshell:

\begin{align}
I^{3d} &\equiv \frac{\la \epsilon f | T^{(coul)} | 3 d \ra \langle \epsilon f \, 3d | V_{ee} | 3d \, \epsilon f \rangle}{E-\epsilon_{f}+\epsilon_{3d}}
\nn \\
I_{tf}^{4s} &\equiv \frac{\la \epsilon' p | T^{(coul)} | 4 s \ra \langle \epsilon f \, 4s | V_{ee} | 3d \, \epsilon' p \rangle}{E -\epsilon_{p}'+\epsilon_{4s}}
\label{3d4scorrelations} \\
I_{tb}^{4s} & \equiv \frac{\la 4 s | T^{(coul)} | \epsilon' p \ra \langle \epsilon f \, \epsilon' p | V_{ee} | 3d \, 4s \rangle}{E +\epsilon_{p}-\epsilon_{4s}}
\nn
\end{align}
We calculated those terms for $E \in [280\, \mathrm{eV}, 380 \, \mathrm{eV}]$ and $\epsilon_{f} =E-279.4 \,  \mathrm{eV}$, where $279.4 \, \mathrm{eV}$ is the Ru-$3d$ ionization energy. $\epsilon_{p}' = \e_f+200 \, eV$ such that the denominator for the intershell contribution is comparable to the real core-hole contribution. The ratio of the two terms is calculated to satisfy $I_{tf}^{4s}/I^{3d} < 0.016$ within the given energy range. The ratio of the backward-propagating term  in the same energy range is $ I_{tb}^{4s}/I^{3d} < 10^{-4}$ due to the large denominator. 

Intrashell correlation of the $\epsilon f-3d$ pair with $\e' d-3 p$ is similarly controlled by the forward and backward terms
\begin{align}
I_{tf}^{3p} & \equiv \frac{\la \epsilon' d | T^{(coul)} | 3 p \ra \langle \epsilon f \, 3p | V_{ee} | 3d \, \epsilon' d \rangle}{E -\epsilon_{d}'+\epsilon_{3p}}
\nn \\
I_{tb}^{3p} & \equiv \frac{\la 3 p | T^{(coul)} | \epsilon' d \ra \langle \epsilon f \, \epsilon' d | V_{ee} | 3d \, 3p \rangle}{E +\epsilon_{d}'-\epsilon_{3p}}\,.
\label{3d3pcorrelations}
\end{align} 
We consider the same range of E  and the corresponding $\epsilon_f$, while $\epsilon_d' = 1\, \mathrm{eV}$.  The respective ratios in the same energy range as above are estimated to be $I_{tf}^{3p}/I < 0.013$ and $I_{tb}^{3p}/I < 1.2 \times 10^{-4}$.  Thus, the overall contribution of terms such (\ref{correlations}) is less than a $1.5\%$. For deeper shells, such as $\mathrm{W}-3d$, the effect of the non-diagonal terms is even less. Those estimates support the intuition that dynamic correlation effects are small in processes involving inner atomic shells, although we have not solved the full RPAE equations of motion. A full calculation within RPAE should be performed for a complete quantitative comparison of RPAE with the relaxed DHF and TDA approximations. 

We further note that the term $I^{3d}$ of Eqs.~(\ref{3d4scorrelations}) diverges close to the ionization threshold $E\to\epsilon_{3d}$. In this limit, the denominator of $I^{3d}$ becomes $\e_f$ leading to a logarithmic divergence upon integration over the continuum energy $\e_f$ in Eq.~(\ref{correlations}). This divergence is removed by resumming the diagrams of Fig.~\ref{fig:MBRPA}, indicating that the inner-shell vacancy must be treated non-perturbatively using relaxed DHF, within the generalized RPAE \cite{AMUSIA1976191}. 

To summarize, the impact of higher-order correlation corrections to the composition analysis of materials in EELS is estimated to be small at least for the most common ionization edges above $\sim O(100\, \mathrm{eV})$.

\bibliography{IonizationCrossSection}

\end{document}